\pdfoutput=1

\documentclass[11pt,twoside,a4paper,cmspaper,final,collab]{cms-tdr}
\def\svnVersion{382113}\def\cmsCernNoTag{CERN-EP/2016-016}\def\cmsCernDate{\today}\def\cmsMessage{Published in the European Physical Journal C as \href{http://dx.doi.org/10.1140/epjc/s10052-016-4504-z}{\doi{10.1140/epjc/s10052-016-4504-z.}}}

\begin{document}\cmsNoteHeader{TOP-12-006}

\hyphenation{had-ron-i-za-tion}
\hyphenation{cal-or-i-me-ter}
\hyphenation{de-vices}
\RCS$Revision: 373480 $
\RCS$HeadURL: svn+ssh://svn.cern.ch/reps/tdr2/papers/TOP-12-006/trunk/TOP-12-006.tex $
\RCS$Id: TOP-12-006.tex 373480 2016-11-10 19:19:27Z alverson $
\newlength\cmsFigWidth
\ifthenelse{\boolean{cms@external}}{\setlength\cmsFigWidth{0.85\columnwidth}}{\setlength\cmsFigWidth{0.4\textwidth}}
\ifthenelse{\boolean{cms@external}}{\providecommand{\cmsLeft}{top}}{\providecommand{\cmsLeft}{left}}
\ifthenelse{\boolean{cms@external}}{\providecommand{\cmsRight}{bottom}}{\providecommand{\cmsRight}{right}}
\ifthenelse{\boolean{cms@external}}{\providecommand{\cmsTable}[1]{\resizebox{\columnwidth}{!}{#1}}}{\providecommand{\cmsTable}[1]{#1}}
\newcommand{\vis}{\ensuremath{\text{vis}}}

\cmsNoteHeader{TOP-12-006}
\title{Measurements of the \texorpdfstring{$\ttbar$}{t-tbar} production cross section in lepton+jets final states in pp collisions at \texorpdfstring{8\TeV}{8 TeV} and ratio of \texorpdfstring{8~to~7\TeV}{8 to 7 TeV} cross sections}

\titlerunning{Measurement of the \ttbar cross section in lepton+jets at 8\TeV}

\date{\today}

\abstract{
A measurement of the top quark pair production ($\ttbar$) cross section in
proton-proton collisions at the centre-of-mass energy of~8\TeV is presented
using data collected with the CMS detector at the LHC,
corresponding to
an integrated luminosity of 19.6\fbinv.  This
analysis is performed in the $\ttbar$ decay channels with one
isolated, high transverse momentum electron or muon and at least four
jets, at least one of which is required to be identified as
originating from hadronization of a b~quark.  The calibration of the
jet energy scale and the efficiency of b~jet identification are
determined from data.  The measured
$\ttbar$ cross section  is $228.5 \pm 3.8\stat \pm 13.7\syst \pm 6.0\lum\unit{pb}$.
This measurement is compared with an analysis of 7\TeV data, corresponding
to an integrated luminosity of 5.0\fbinv,
to determine the ratio of 8\TeV to 7\TeV cross sections, which is found to be
$1.43 \pm 0.04\stat \pm 0.07\syst \pm 0.05\lum $.
The measurements are in agreement with QCD predictions up to
next-to-next-to-leading order.  }

\hypersetup{%
pdfauthor={CMS Collaboration},%
pdftitle={Measurements of the t-tbar production cross section in lepton+jets final states in pp collisions at 8 TeV and ratio of 8 to 7 TeV cross sections},%
pdfsubject={CMS},%
pdfkeywords={CMS, LHC, physics, top quark, cross section}}

\maketitle

\section{Introduction}

Top quarks are abundantly produced at the CERN LHC.
The predicted
top quark pair production cross section ($\sigma_{\ttbar}$) in
proton-proton (pp) collisions, at a centre-of-mass energy
of $8\TeV$, is 253\unit{pb}, with theoretical
uncertainties at the level of 5--6\%.
A precise measurement of $\sigma_{\ttbar}$
is  an important
test of perturbative quantum chromodynamics (QCD) at high energies.
Furthermore,
precision $\ttbar$ cross section measurements can be used to constrain
the top quark mass $m_{\PQt}$ and QCD
parameters,
such as the strong coupling
constant $\alpha_S$~\cite{top-12-022},
or the parton distribution functions (PDF)
of the proton~\cite{mitov2}.

The \ttbar production cross section
was measured at the LHC
at $\sqrt{s} = 7$ and 8\TeV~\cite{top-10-001, top-11-002, top-10-002, top-10-003, top-11-005, top-11-003, top-11-007,
top-11-004, top-11-006, atlas-2011, atlas-2013tau, atlas-2012l, atlas-2012ltau,
atlas-2012dilep1, atlas-2012dilep2, atlas-2014dilep,
top-12-007, top-12-026, top-14-018, atlas-2014dilep, atlas-2015,
atlas-2015ljets, atlas-2015-br, top-13-004}.
In this paper, a
measurement of the \ttbar production cross section
in the final state with one high transverse momentum lepton (muon or electron) and jets
is presented using the 2012 data set at  $\sqrt{s} = 8\TeV$,
collected by the CMS experiment at the LHC
and corresponding to an integrated luminosity of
19.6\fbinv.
To measure the cross section ratio,
where several systematic uncertainties cancel,
the 2011 data set at $\sqrt{s} = 7\TeV$,
corresponding to an integrated luminosity of 5.0\fbinv,
has been concurrently analyzed
with a similar strategy
to the one developed for the cross section measurement at 8\TeV.
The
new measurement agrees very well with the previously published
CMS result~\cite{top-11-003}.
The larger statistical uncertainty of the present measurement
with respect to the previous one is due
to the simultaneous determination of the \cPqb~tagging efficiency,
as discussed in Section~\ref{sec:syst}.
Similarly to the 8\TeV analysis, an additional signal modelling uncertainty
has been considered in the 7\TeV analysis, as reported in
Section~\ref{sec:syst}.

In the standard model,
top quarks are predominantly
produced in pairs
via the strong interaction
and decay almost exclusively into a $\PW$~boson
and a \cPqb~quark.
The event signature is determined by the
subsequent decays of the two $\PW$~bosons.
This analysis uses  lepton+jets decays into
muons or electrons, where one of the $\PW$~bosons decays into
two quarks and the other
to a lepton and a neutrino.
Decays of the $\PW$ boson  into a tau lepton and a neutrino can enter
the selection if the tau lepton decays leptonically.
The top quark
decaying into a \cPqb~quark and a leptonically decaying $\PW$~boson
is defined in the following as the ``leptonic top quark'', while
the other top quark
is referred to as ``hadronic top quark''.
For the {\ttbar} signal two  jets
result from the hadronization of the $\cPqb$  and
$\cPaqb$ quarks (\cPqb~jets),
thus $\cPqb$ tagging algorithms are employed for
the identification
of $\cPqb$ jets
in order to improve the purity of the \ttbar candidate sample.

The technique for extracting the \ttbar cross section
consists of a binned log-likelihood fit of signal and
background to the distribution of a
discriminant variable in data showing a good separation between signal and background: the invariant mass of
the $\cPqb$~jet related to the leptonic top quark and the
lepton $\ell$ ($M_{\ell\cPqb}$).
The mass of the three-jet combination with the
highest transverse momentum
in the event ($M_3$) is used as a discriminant
in an alternative analysis.
The $M_{\ell\cPqb}$ variable is related to the leptonic
top quark mass, while $M_3$  is a measure for the
hadronic top quark mass.
Both quantities
provide a good separation
between signal and background processes.

The analysis employs calibration techniques  to reduce
the experimental uncertainties related to \cPqb~tagging
efficiencies
and jet energy scale (JES).
The $\ttbar$ topology is reconstructed using a jet
sorting algorithm in which the \cPqb~jet most likely originating from the
leptonic top quark
is identified. The \cPqb~tagging
efficiency is then determined from a \cPqb-enriched sample,
in the peak region of the $M_{\ell\cPqb}$ distribution,
correcting for the contamination from
non-\cPqb~jets, following the method described in
Refs.~\cite{CMS_btag04,CMS_btag04_bis}.
The rate of jets that are wrongly tagged as originating from a \cPqb~quark
is also measured using data as described in~\cite{maes}.
Independently, the JES
is determined using the jets associated with the
hadronically decaying $\PW$~boson
by correcting the reconstructed mass of the $\PW$~boson in the simulation
to that determined from the data.

The results of the cross section measurements are given both for
the visible region,
\ie for the phase space corresponding to the event selection,
and for the full phase space. The visible region is defined by requiring
the presence in the simulation
of exactly one lepton, one neutrino, and at least
four jets
passing the selection
criteria, as presented in Section~\ref{sec:xsec}.

This paper is structured as follows: after a description
of the CMS detector (see Section~\ref{sec:cms}),
the data and the simulated samples are
discussed in Section~\ref{sec:MC}, while Section~\ref{sec:selection}
is dedicated to the event selection.
The analysis technique
and the  impact of the systematic uncertainties
are addressed in Section~\ref{sec:xsec} and in Section~\ref{sec:syst}.
The results of the cross section
measurements
are discussed in Section~\ref{sec:comb}.
Section~\ref{sec:M3}
describes the alternative
analysis based on $M_3$,
followed by a summary
in Section~\ref{sec:summary}.

\section{The CMS detector}
\label{sec:cms}

The central feature of the CMS
apparatus is a superconducting solenoid,
of 6~m internal diameter, providing an axial magnetic field of 3.8\unit{T}.
Within the solenoidal field volume are a silicon pixel and strip tracker
which measure charged particle trajectories in the pseudorapidity range
$\abs{\eta} < 2.5$.
Also within the field volume, the silicon detectors are surrounded by
a lead tungstate crystal electromagnetic calorimeter ($\abs{\eta}<3.0$) and
a brass and scintillator hadron
calorimeter ($\abs{\eta}<5.0$) that provide high-resolution energy and
direction measurements of electrons and hadronic jets.
Muons are measured in gas-ionization detectors embedded in the steel
magnetic flux-return yoke outside the solenoid.
The muon detection systems
provide muon
detection in the range $\abs{\eta} < 2.4$.
A two-level trigger system selects the
pp collision
events for use in physics analysis.
A more detailed description of the
CMS detector, together with a definition
of the coordinate system
used and the relevant kinematic variables,
can be found elsewhere~\cite{cmsdet}.

\section{Data and simulation}
\label{sec:MC}

The cross section measurement is performed using the 8\TeV
pp collisions recorded by the CMS experiment in 2012,
corresponding to an integrated luminosity of
$19.6 \pm 0.5\fbinv$~\cite{lumi_2012},
and the
2011 data set at $\sqrt{s} = 7\TeV$, corresponding to
an integrated luminosity of
$5.0 \pm 0.2\fbinv$~\cite{lumi_2011}.

The  \ttbar events are simulated
using the Monte Carlo (MC) event generators
\MADGRAPH (version 5.1.1.0)~\cite{MG5,MG5bis}
and \POWHEG (v1.0 r1380)~\cite{powheg1,powheg2}.
In \MADGRAPH the top quark pairs are generated at leading order
with up to three additional high-$\pt$ jets.
The \POWHEG generator implements matrix elements
to next-to-leading order (NLO) in perturbative QCD,
with up to one additional jet.
The mass of the top quark is set to
$172.5\GeV$.
The CT10~\cite{gao} PDF set is used by \POWHEG and the
CTEQ6M~\cite{Alekhin,Botje,ball}
by \MADGRAPH.
The \PYTHIA (v.6.426)~\cite{Sjostrand:2006za}
and \HERWIG (v.6.520)~\cite{herwig} generators
are used to model the parton showering.
The \PYTHIA shower matching is done using the MLM
prescription~\cite{mlm, Alwall09}.

{\sloppy
The top quark pair production cross section values are predicted to be
$177.3^{+4.6}_{-6.0}~\text{(scale)} \pm 9.0~(\mathrm{PDF}{+}\alpha_S)\unit{pb}$ at 7\TeV and
$252.9^{+6.4}_{-8.6}~\text{(scale)} \pm 11.7~(\mathrm{PDF}{+}\alpha_S)\unit{pb}$ at 8\TeV,
as calculated with the {\sc Top++ 2.0} program to next-to-next-to-leading order
(NNLO) in perturbative QCD,
including soft-gluon resummation to next-to-next-to-leading
logarithmic (NNLL) order (Ref.~\cite{topplusplus} and references therein),
and assuming
$m_{\PQt} = 172.5\GeV$.
The first uncertainty comes from the independent variation
of the factorization and renormalization scales, while the second one
is associated to variations in the PDF and $\alpha_S$
following the PDF4LHC prescription with the MSTW2008 68\% confidence
level NNLO,
CT10 NNLO, and NNPDF2.3 5f FFN PDF sets
(Refs.~\cite{Alekhin,Botje} and references
therein, and Refs.~\cite{gao,ball}).
\par}

The top quark transverse momentum is reweighted
in samples simulated with \MADGRAPH and \POWHEG,
when interfaced to
\PYTHIA,
in order
to better describe the
\pt distribution observed in the data.
Based on studies of differential
distributions~\cite{top-11-013,top-12-028}
in the top quark transverse momentum,
an event weight  $w = \sqrt{\smash[b]{w_1 \, w_2}}$ is applied,
where the  weights $w_i$ of the two top quarks are given
as a function of the
generated top quark $\pt$ values:
$w_i = \exp(0.199 - 0.00166 \ \pt^i/{\GeVns})$ at 7\TeV, and
$w_i = \exp(0.156 - 0.00137 \ \pt^i/{\GeVns})$ at 8\TeV.
This reweighting is only applied to the phase space corresponding
to the experimental selections in the muon and electron channels.
The agreement between data and samples generated with \POWHEG interfaced
with \HERWIG is found to be satisfactory, and no reweighting is applied
in this case.

The  $\PW/\PZ$+jets events,
\ie the associated production of $\PW/\PZ$ vector bosons with jets,
with leptonic decays of the $\PW/\PZ$
bosons, constitute
the largest background. These are also simulated using \MADGRAPH
with matrix elements corresponding to at least one jet  and up to four jets.
The $\PW/\PZ$+jets events are generated inclusively with respect to the
jet flavour.
Drell--Yan production of charged leptons is generated for dilepton
invariant masses above 50\GeV, as those events constitute the
relevant background in the phase space of this analysis.
The contribution from Drell--Yan events with dilepton invariant
masses below  50\GeV is negligible, as verified with a sample
with a mass range of 10--50\GeV.
Single top quark production is simulated with
\POWHEG.
The background processes are normalized to NLO and NNLO
cross section calculations~\cite{fewz1,fewz2,mcfm1,mcfm3,mcfm4},
with the exception of the QCD multijet background,
for which the normalization is obtained from data in the $M_3$ analysis
(see Section~\ref{sec:M3}). In the $M_{\ell\cPqb}$ analysis the multijet
background is reduced to a negligible fraction (see Section~\ref{sec:selection})
and thus not considered further.

Pileup signals, \ie extra activity
due to additional
pp interactions in the same bunch crossing,
are incorporated by simulating additional interactions
with a multiplicity matching the one
inferred from data.
The CMS detector response is
modeled
using
\textsc{Geant4}~\cite{Geant4}. The simulated events are
processed by the same reconstruction software as the collision data.

\section{Reconstruction and event selection}
\label{sec:selection}

This analysis focuses
on the selection of \ttbar lepton+jets
decays in the muon and electron channels, with similar
selection requirements applied for the two channels.
Muons, electrons, photons, and neutral and charged hadrons are reconstructed and
identified by
the CMS particle-flow (PF)
algorithm~\cite{ipartf09,ipartf10}.
The energy of muons is obtained from the
corresponding track momentum using the combined information of the silicon tracker and
the muon system~\cite{CMS_mu}.
The energy of electrons is determined from a combination of the track momentum
in the tracker, the corresponding cluster energy in the
electromagnetic calorimeter,
and the energy sum of
all bremsstrahlung photons associated to the track~\cite{CMS_ele_new}.
The vertex with the largest $\pt^2$ sum of the tracks associated to it
is chosen as primary vertex.

Candidate \ttbar events are first accepted by
dedicated triggers
requiring at least one muon or electron.
Lepton isolation requirements are applied to improve the purity
of the selected sample.
At the trigger level the relative muon isolation, the sum of transverse momenta of other particles
in a cone of size  $\Delta R =\sqrt{\smash[b]{(\Delta\phi)^2+(\Delta\eta)^2} }  = 0.4$
around the direction of the candidate muon divided by
the muon transverse momentum, is required to be
less than 0.2.
Similarly, for electrons, the corresponding requirement is
less than 0.3 in a cone of size 0.3.
Events with a muon in the final state
are triggered on the presence of a muon candidate with
$\pt > 24\GeV$ and
$\abs{\eta} <2.1$.
Events with an electron candidate with $\abs{\eta} < 2.5$ are accepted by
triggers requiring an
electron with $\pt > 27\GeV$.

Tighter
$\pt$ requirements
are applied in the offline selections.
Muons are required to have a good quality~\cite{CMS_mu}
track with
$\pt > 25 \GeV$
and
$\abs{\eta} <2.1$.
Electrons are identified
using a combination of
the shower shape information and track-electromagnetic cluster
matching~\cite{CMS_ele_new},
and are required to have $\pt\ > 32\GeV$
and $\abs{\eta} < 2.5$,
with the exclusion of the transition region between the barrel and
endcap electromagnetic
calorimeter, $1.44 < \abs{\eta} < 1.57$.
Electrons identified to come from photon conversions~\cite{CMS_ele_new} are vetoed.
Correction factors for
trigger and lepton identification efficiencies
have been determined with a tag-and-probe method~\cite{tag_probe}
from data/simulation comparison
as a function of the lepton $\pt$ and $\eta$,
and are applied to the simulation.

Signal events are required to have at least one pp
interaction vertex,
successfully reconstructed from at least four tracks,
within limits on the longitudinal and radial coordinates~\cite{CMS_vertex},
and
exactly one muon, or electron, with an
origin consistent with the reconstructed vertex
within limits on the impact parameters.
Since the lepton from the $\PW$~boson decay is expected to be isolated from
other activity
in the event, isolation requirements are applied.
A relative isolation is defined as
$I_{\text{rel}}
=(I_{\text{charged}} + I_{\text{photon}} + I_{\text{neutral}}) / \pt$, where
\pt is the transverse momentum of the lepton and $I_{\text{charged}}$,
$I_{\text{photon}}$, and $I_{\text{neutral}}$ are the sums of the
transverse energies of the charged particles, the
photons, and the
neutral particles not identified as photons,
in a cone   $\Delta R < 0.4\,(0.3)$ for muons (electrons)
around the
lepton direction, excluding the lepton itself.
The relative isolation $I_{\text{rel}}$ is required to be
less than 0.12 for muons and 0.10
for electrons.
Events with more than one lepton
candidate with relaxed requirements are vetoed
in order to reject $\PZ$~boson or \ttbar decays into dileptons.

The missing  energy in the transverse plane (\ETmiss)
is defined as the magnitude
of the projection on the plane perpendicular to the beams of the vector
sum of the momenta of all PF candidates.
It is required to be larger than 30\GeV in
the muon channel and larger than 40\GeV in the electron channel,
because of the larger multijet background.

Jets are clustered from the
charged and neutral particles
reconstructed with the PF algorithm,
using the anti-$\kt$ jet algorithm \cite{ktalg}
with a radius parameter of 0.5.
Particles identified as isolated muons or electrons
are not used in the jet clustering.
Jet energies are corrected for nonlinearities due to different
responses in the calorimeters and for the
differences between measured
and simulated responses~\cite{jec_11}.
Furthermore, to account for extra activity
within a jet cone
due to
pileup,
jet energies are corrected~\cite{ipartf09,ipartf10}
for charged hadrons that belong
to a vertex other than the  primary vertex,
and for the amount of pileup expected
in the jet area from neutral jet constituents.

At least four jets are required
with
$\pt > 40\GeV$
and $\abs{\eta} < 2.5$.
An additional global calibration factor of the jet energy scale is
obtained by fitting the $\PW$~boson mass distribution in the data and in
the simulation.  The scale factor is determined as the ratio of the
$\PW$~boson mass reconstructed from non-\PQb-tagged jet pairs in data and in the
simulation.  This scale correction is applied in the simulation to all
jets before the selection requirements are implemented. It
largely reduces the systematic uncertainty related to the jet energy scale,
discussed in Section~\ref{sec:syst}.

\begin{figure*}[tbh]
\centering
\includegraphics[width=0.48\textwidth]{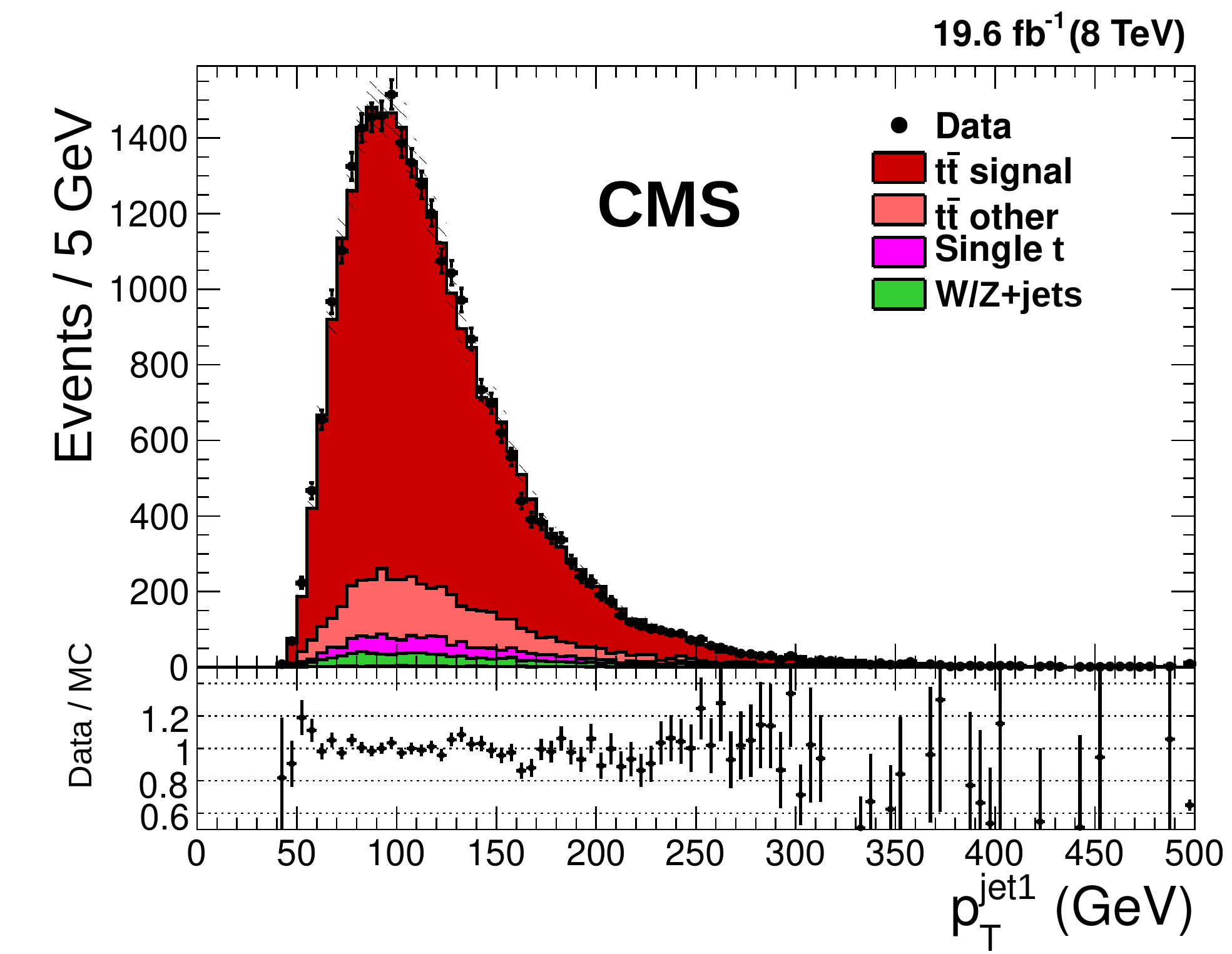}
\includegraphics[width=0.48\textwidth]{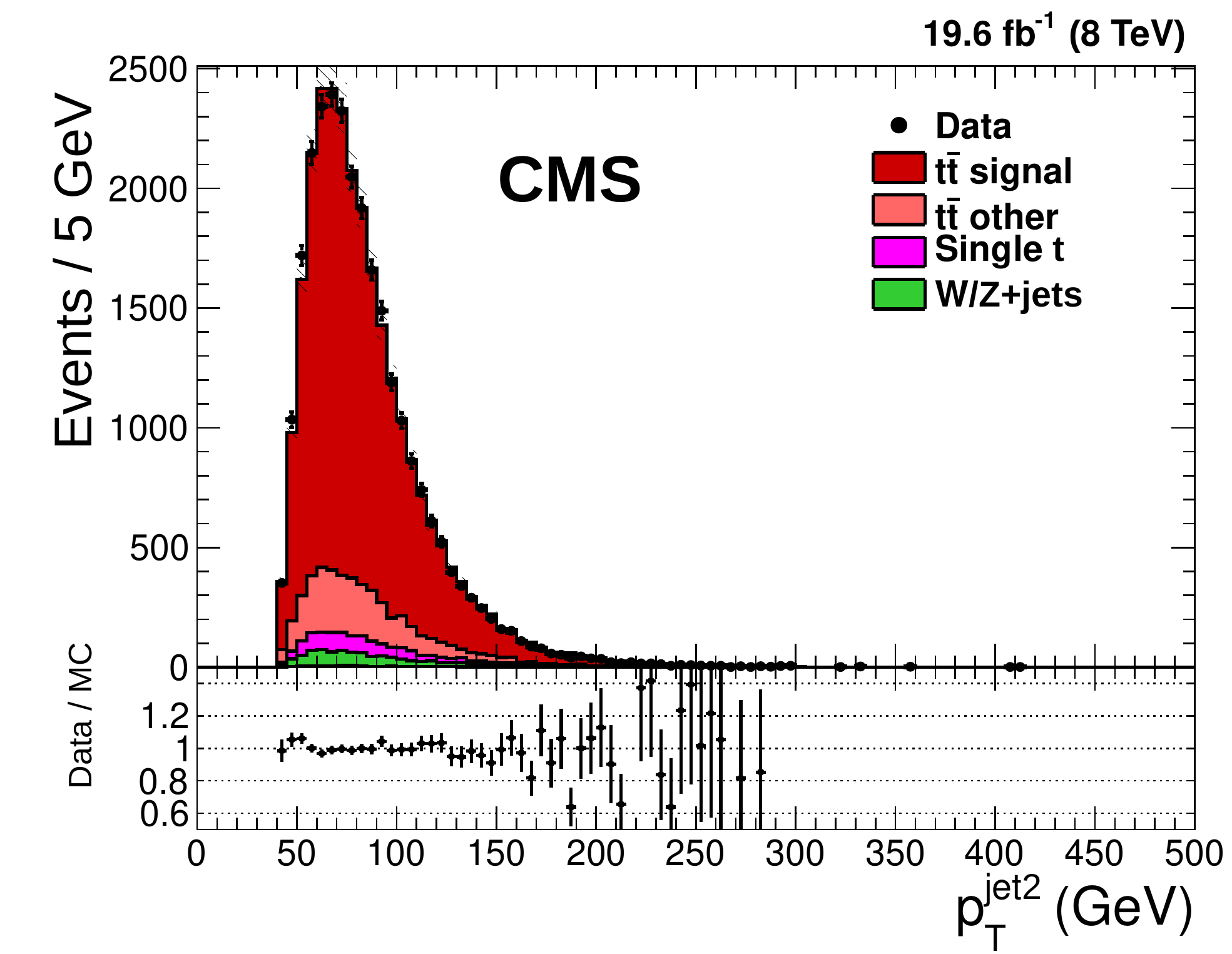} \\
\includegraphics[width=0.48\textwidth]{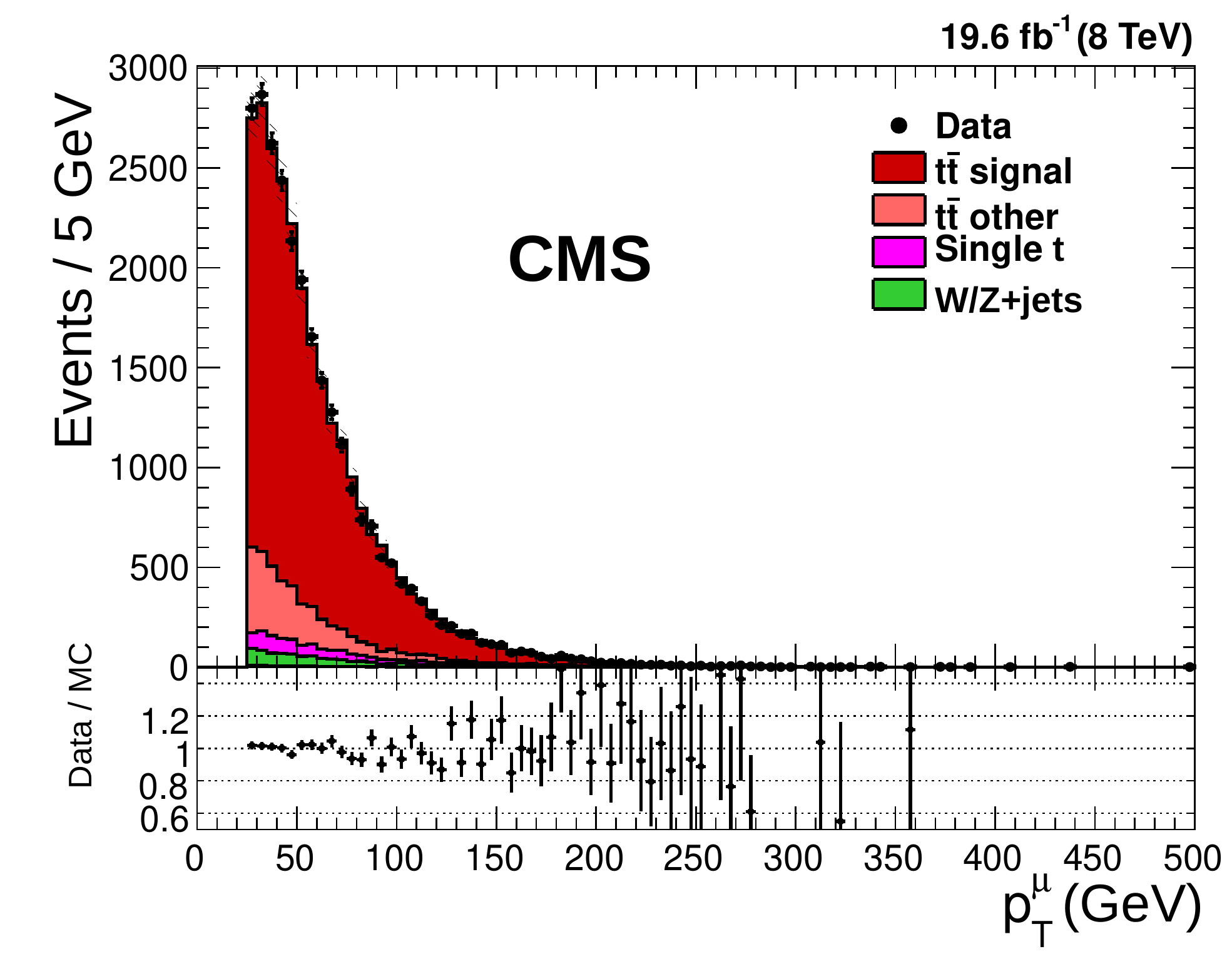}
\includegraphics[width=0.48\textwidth]{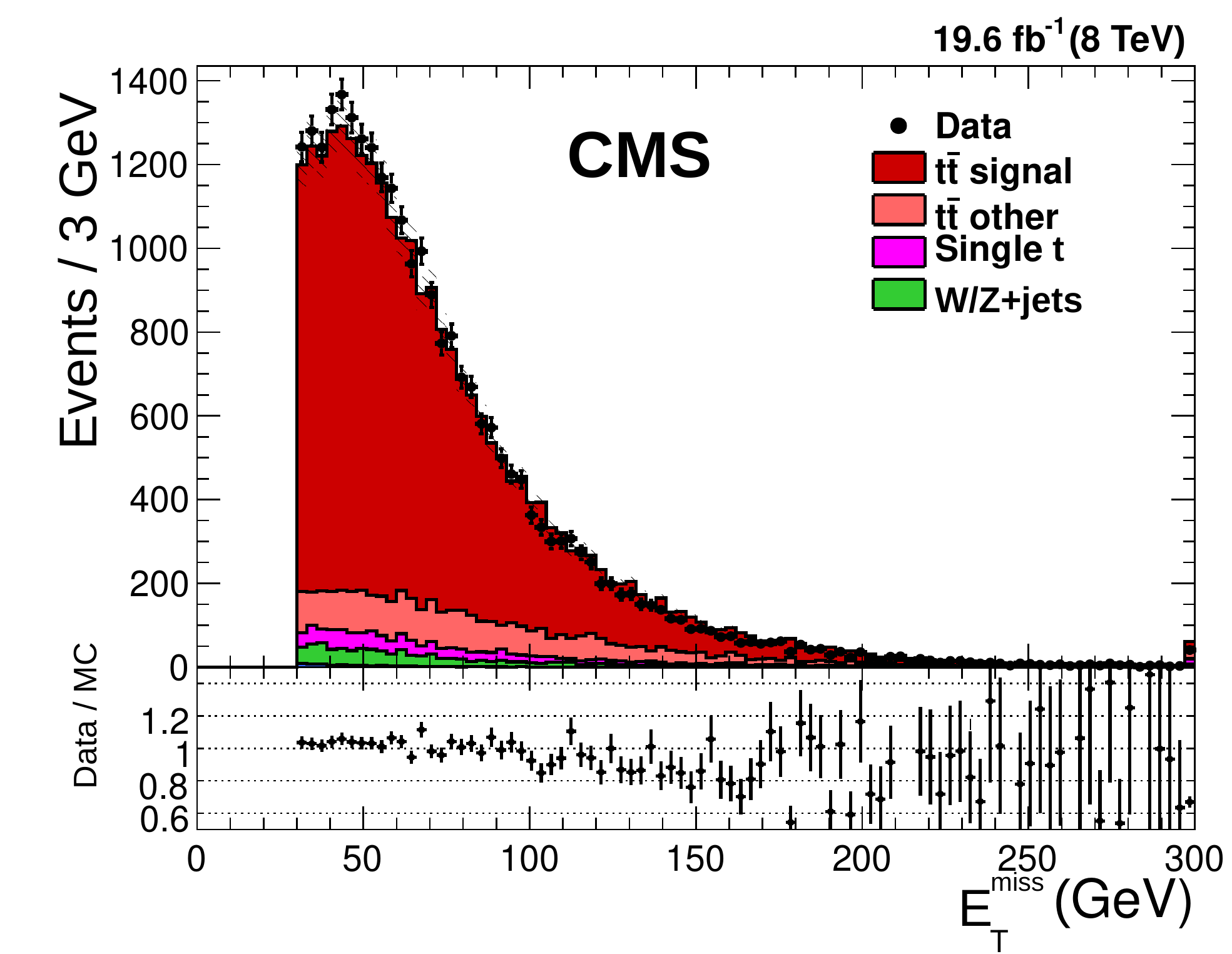}
\caption{
Transverse momentum distributions
of the first- and second-leading jet (top),
the muon and \ETmiss distribution
(bottom) for all relevant processes
in the muon+jets channel
with the requirement of at least one \cPqb-tagged jet.
The simulation is normalized to the standard model cross section values
and \pt-reweighting is applied to the \ttbar contribution.
The multijet background is negligible and not shown.
The distributions are already corrected for the
\cPqb~tagging efficiency scale factor.
The hashed area shows the uncertainty in the luminosity
measurement and the \cPqb~tagging systematic uncertainty.
The last bin includes the overflow.
The ratio between data and simulation is shown in the lower panels
for bins with non-zero entries.
}
\label{fig:btag_example}
\end{figure*}

To reduce contamination from background processes,
at least one of the jets has to be identified as a \cPqb~jet.
The \cPqb~tagging algorithm used is the
``combined secondary vertex'' (CSV) algorithm at the
medium working point~\cite{CMS_btag04,CMS_btag04_bis},
corresponding to a misidentification probability of about 1\%
for light-parton jets (mistag
rate) and an efficiency for \cPqb~jets in the range 60--70\%
depending
on the jet {\pt} and pseudorapidity.
Figure~\ref{fig:btag_example} shows
kinematic distributions
after applying the \cPqb~tagging requirement.
Good agreement between data and simulation is observed.

{\tolerance=5000
The  $M_{\ell\cPqb}$ analysis uses control samples in data for the estimation
of the b tagging efficiency, as described
in Refs.~\cite{CMS_btag04, CMS_btag04_bis, maes}.
Among the four leading jets, three are assigned to
the hadronically decaying top quark through a $\chi^2$ sorting
algorithm using top quark and $\PW$~boson mass constraints.
The remaining fourth jet
is the \cPqb~jet candidate assigned to the leptonically decaying top quark.
The \cPqb~tagging algorithm is only applied to this \cPqb~jet candidate.
}

Owing to differences in the triggers and in the centre-of-mass energies,
in the 7\TeV analysis
slightly different
selection criteria are applied
on the lepton $\pt$ and \ETmiss.
The muon transverse momentum is required to be larger than 26\GeV, while
the electron $\pt  $ has to be larger than 30\GeV.
No explicit \ETmiss
requirement
is needed in the muon channel.
Events
with $\ETmiss  > 30\GeV$
are selected in the electron channel.

\section{Visible and total cross section measurements }
\label{sec:xsec}

The number of \ttbar events is determined
with a binned maximum-likelihood fit of distributions
(templates), describing signal and background processes, to the data
sample passing the final selection, by fitting $M_{\ell\cPqb}$,
the invariant mass distribution of the \cPqb~jet and the lepton.

The \ttbar visible ($\sigma_{\ttbar}^{\vis}$) and total ($\sigma_{\ttbar}$) production cross sections
are extracted from the number of \ttbar events observed in the data using the equations
\begin{equation} \label{eq:cross_section}
\sigma_{\ttbar}^{\vis} = \frac{N_{\ttbar}}{ L  \, \varepsilon_{\ttbar}}  ,
\hspace{2.0cm}  \sigma_{\ttbar} = \frac{\sigma_{\ttbar}^{\vis}}{A} ,
\end{equation}
where $N_{\ttbar}$ is the number of \ttbar events
(including both signal events from the lepton+jets channel
considered and events from other decay channels) extracted from the fit,
$L$ is the integrated luminosity,  $A$ is the
$\ttbar$ acceptance,
and
$\varepsilon_{\ttbar}$ is
the $\ttbar$ selection efficiency
within the acceptance requirements outlined in the next section.
Results are presented for both
the visible and total cross section,
in order to separate experimental uncertainties
from theoretical assumptions as much as possible.

\begin{figure*}[htb]
  \centering
\includegraphics[width=0.40\textwidth]{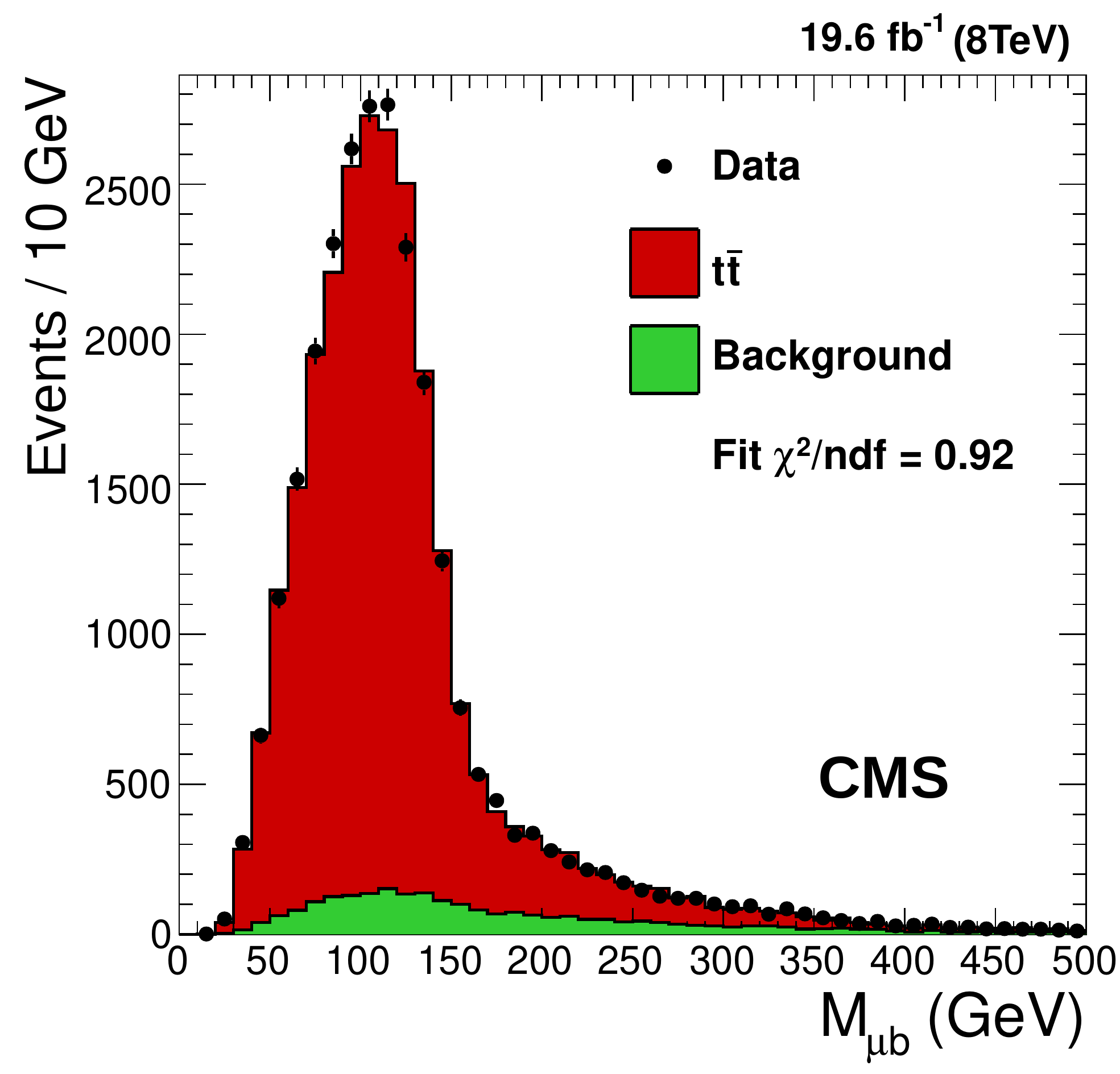}
\includegraphics[width=0.40\textwidth]{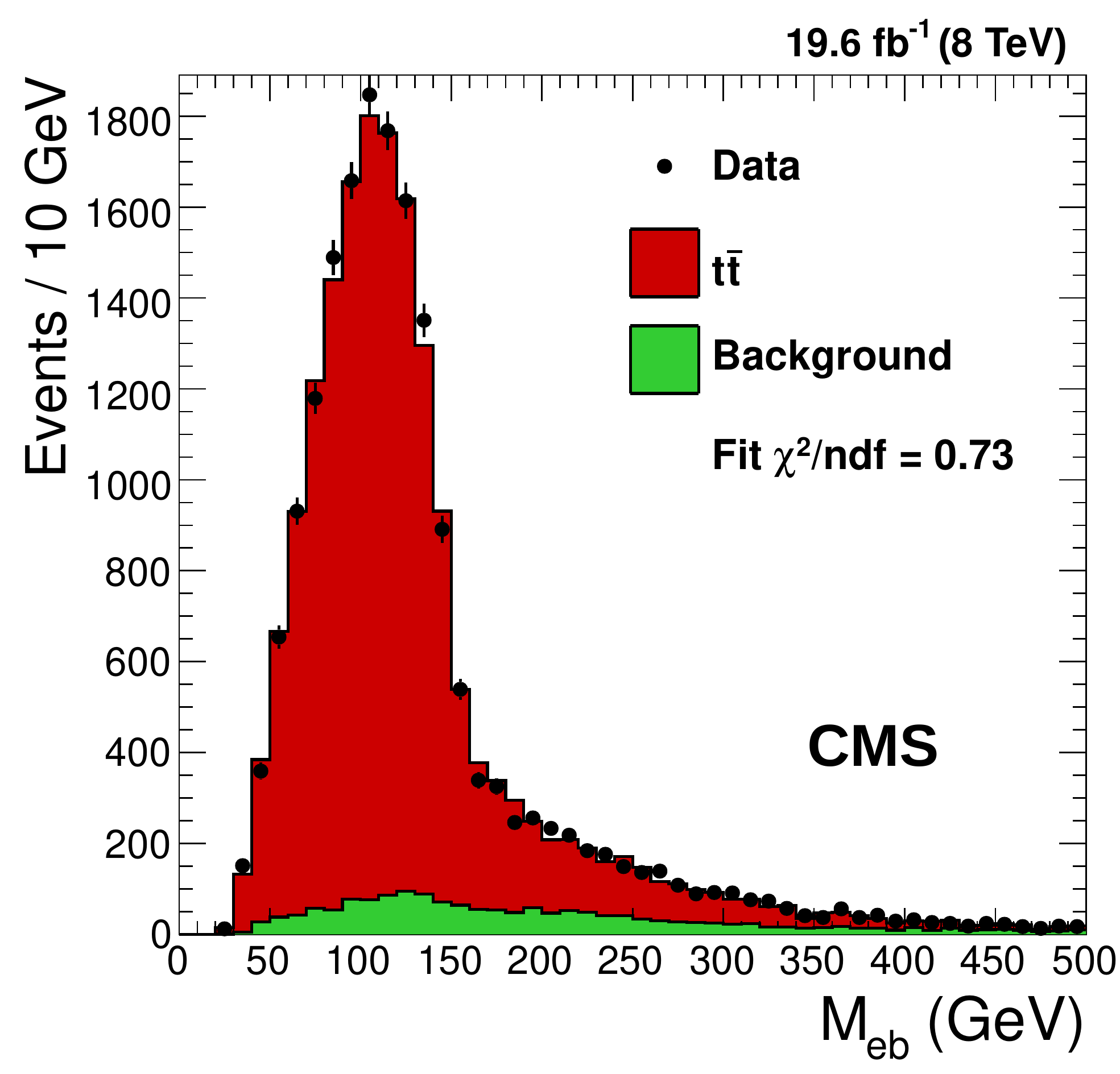}
 \caption{
Distributions of the lepton-jet mass in the muon+jets (left)
and electron+jets (right) channels,
rescaled
to the fit results.
}
\label{fig:mlfit_res}
\end{figure*}

One template is used for \ttbar events (both for the \ttbar
signal events and the other \ttbar events passing the selection criteria)
and one template for all background processes ($\PW/\PZ$+jets and
single top quark production). The fit is performed in the range 0--500\GeV.
Figure~\ref{fig:mlfit_res} shows the results for the fit to the
data distributions in the muon and electron channels.

\subsection{Acceptance}

The \ttbar
acceptance $A$ corresponding to the visible phase space depends on the
theoretical model and it is
determined at the generator level
by requiring
the presence of exactly one lepton, one neutrino, and
at least four jets, passing
$\pt$  and $\abs{\eta}$ selection criteria
similar to the ones delineated in Section~\ref{sec:selection}.
For simplicity a single acceptance
definition, corresponding to the tightest selection criteria,
is used for both channels
at each centre-of-mass
energy:
exactly one muon, or electron, with
$\pt > 32\GeV$  and $\abs{\eta} < 2.1 $,
one neutrino with $\pt > 40\GeV$,
and at least four jets with $\pt > 40\GeV$  and $\abs{\eta} < 2.5 $.

The acceptance values include contributions from other
\ttbar
decay channels, in particular from the dilepton channel, at the
level of about 9\%.

The acceptance values are provided in Table~\ref{tab:acceptance}
for the two generators used in this analysis, \MADGRAPH and
\POWHEG.
The acceptance values are in agreement
at the 1--2\% level at 8\TeV and at better than 5\% at 7\TeV.
This different level of agreement is due to the fact that the
common
acceptance definition described above corresponds
the tightest $\pt$ criteria, \ie to the $\pt$ requirements
of the electron channel at $\sqrt{s} = 8$\TeV.
The reweighted acceptance is determined as the number of reweighted
\ttbar events in the visible phase space, \ie the sum of the
weights, divided by the total number of (non-reweighted) \ttbar
events.

The statistical uncertainty in the acceptance calculations is below 3\%. The
theoretical systematic uncertainties evaluated by varying
the PDFs (Section~\ref{sec:syst})
or the matching thresholds
are in the range 0.1--0.2\%. Variation of the factorization and renormalization scale
induces a variation of up to 2\% in the acceptance. These variations are
already
included in the systematic uncertainties quoted in
Section~\ref{sec:syst}.

In the following, top quark $\pt$-reweighting~\cite{top-11-013,top-12-028}
is always applied to the visible phase space
as it provides a better agreement between data
and simulation.
On the other hand, given that the event weights were only determined
in the phase space corresponding to the experimental selection,
they have not been used for the extrapolation to the total cross
section. Therefore, the non-reweighted
acceptance is used to determine the total cross section. However,
rescaling by the ratio of the values provided in Table~\ref{tab:acceptance}
would allow a determination of the total cross
section with the reweighted acceptance.
The visible cross section does not depend on the acceptance~$A$.

\begin{table*}[htp]
\topcaption{Average acceptance values for the muon and electron channels
obtained with \MADGRAPH and \POWHEG
at $\sqrt s = 7$~and~8\TeV,
without and with top quark $\pt$-reweighting applied.
The statistical uncertainty
is 0.0004, \ie below 3\%.
The theoretical uncertainties are at the level of 2\%, as discussed in
the text.
}
\centering
\renewcommand{\arraystretch}{1.2}
\begin{tabular}{ccccc}
 \hline
 \multirow{2}{*}{Generator} & \multicolumn{2}{c}{$A$ ($\sqrt s = 7\TeV$)} & \multicolumn{2}{c}{$A$
($\sqrt s = 8\TeV$) }\\
 & no rew. & with rew. & no rew. & with rew. \\
\hline
\MADGRAPH & 0.0158 & 0.0156 & 0.0166 & 0.0162 \\
\POWHEG & 0.0151 & 0.0149 & 0.0163 & 0.0161 \\
\hline
\end{tabular}
\label{tab:acceptance}
\end{table*}%

\subsection{Selection efficiency}

The
selection efficiency within the
acceptance,
$\varepsilon_{\ttbar}$, is reported in Table \ref{tab:sel_eff_signal}.
It is determined from the $\pt$-reweighted \MADGRAPH simulated sample
as the number of
events passing the selection criteria outlined in Section~\ref{sec:selection},
over the number of events passing the
acceptance requirements defined above.
The selection efficiency includes the effects of trigger requirements,
lepton and jet
identification criteria, and \cPqb~tagging efficiency, which is directly
determined from data.
A signal selection efficiency within acceptance
of
$32\% $ in the muon channel and $21\% $ in the electron channel is determined.
Similar values (37\% and 22\%, respectively)
are obtained at $\sqrt s = 7\TeV$.
For the muon channel the common acceptance requirements used for both
channels are tighter than
the selection requirements, thus the muon
channel efficiency is significantly larger than the
electron channel efficiency.
The \ttbar selection efficiency,
$A \varepsilon_{\ttbar}  $, is the number of
selected \ttbar events out of all produced \ttbar pairs, in all
decay channels.

\begin{table*}[tbh]
  \topcaption{Signal selection efficiencies, at $\sqrt s = 8\TeV$,
  determined from simulation using \MADGRAPH.
The non-reweighted acceptance from
Table \ref{tab:acceptance} is used. The relative statistical
uncertainty on $\varepsilon_{\ttbar}$ is below 3\%.
}
    \label{tab:sel_eff_signal}

  \centering
\renewcommand{\arraystretch}{1.2}
    \begin{tabular}{ccccc}
      \hline
        Channel     & $\varepsilon_{\ttbar}$ ($\sqrt s = 7\TeV$)  & $ A \varepsilon_{\ttbar}$ ($\sqrt s = 7\TeV$)
                    & $ \varepsilon_{\ttbar}$ ($\sqrt s = 8\TeV$)  & $ A \varepsilon_{\ttbar}$ ($\sqrt s = 8\TeV$)    \\
      \hline
$\mu$+jets     & $ 37\% $ & $ 0.58\%$  & $ 32\% $  & $ 0.53\%$  \\
  $\Pe$+jets       & $ 22\% $ & $ 0.36\%$   & $ 21\% $  & $ 0.35\%$  \\
      \hline
    \end{tabular}
\end{table*}

\section{Systematic uncertainties}
\label{sec:syst}

Systematic uncertainties are determined
by varying each source within its estimated uncertainty and
by propagating the variation to the cross section measurements.
Template shapes and signal efficiencies are varied together
according to the systematic uncertainty considered.
The uncertainty is given by the shift in the fitted cross section
and is cross-checked by repeating its estimation with pseudo-experiments
using
simulation. The systematically varied template shapes
are fit to pseudo-data generated using the nominal template shapes
and normalizations. The validation with pseudo-experiments shows
that the fit performs as expected.
All systematic uncertainties, except the ones related to
\cPqb~tagging and to the estimation of the multijet background,
 are common
to both the $M_{\ell\cPqb}$  and the $M_3$ measurements.

The effect of uncertainties in the JES
is evaluated
by varying the JES within the $\pt$- and $\eta$-dependent
uncertainties given in Ref.~\cite{jec_11}.
The final JES of the simulation is matched to that in data
by applying an additional global correction factor $\alpha$ to all jet momenta before
selection. The $\alpha$ calibration values are individually determined for nominal
conditions and for each of the variations related to JES and JER.
In addition to the
selection described in Section~\ref{sec:selection}, two \cPqb-tagged jets
are required in order to increase the signal purity. The mass of the hadronically
decaying $\PW$~boson is reconstructed as the dijet invariant mass
from all combinations of
non \cPqb-tagged jets.
The dijet invariant mass distributions are fitted in data and in
simulation with a function describing the $\PW$~boson signal peak and the
dijet combinatorial background.  The $\alpha$ values are
determined as the ratios of the fitted $\PW$~boson masses in data and in simulation.
In the $M_{\ell\cPqb}$ analysis $\alpha = 1.011 \pm 0.004 $ is obtained with the nominal samples
both in the muon and electron channels, with variations of the order of $\pm$1.5\%
for the samples with down and up variations of the JES. The same values
are determined by the $M_3$ analysis.
This additional calibration reduces the size of the JES
systematic uncertainty by approximately 60\%.
The JES uncertainty, reported in Table~\ref{tab:SYST_JES},
consists of several sources, all propagated individually.
Details of the individual contributions are explained in~\cite{jme-14-003}.

\begin{table}[hbp]
\centering
\topcaption{Components (in \%) of the JES uncertainty at 8\TeV in the muon
and electron channels.
The correlation coefficients used in their combination are also shown.}
\label{tab:SYST_JES}
\renewcommand{\arraystretch}{1.2}
\cmsTable{
\begin{tabular}{lccc}       \hline
Source  & $\mu$+jets & $\Pe$+jets & Correlation  \\
\hline
Absolute scale                          & ${\pm}0.33$ & ${\pm}0.40$ & 0.0 \\
Global jet scale factor $\alpha$          & ${\pm}0.59$ & ${\pm}0.39$ & 0.0  \\
Relative FSR                            & ${\pm}0.46$ & ${\pm}0.41$ & 1.0   \\
Relative \pt                            & ${\pm}0.67$ & ${\pm}0.57$ & 1.0  \\
Flavour JES                             & ${\pm}1.84$ & ${\pm}1.79$ & 1.0  \\
Flavour JES fragmentation               & ${\pm}0.50$ & ${\pm}0.46$ & 1.0   \\
Flavour JES semileptonic BR             & ${\pm}0.11$ & ${\pm}0.16$ & 1.0 \\
High-\pt extra                          & ${\pm}0.18$ & ${\pm}0.23$ & 1.0  \\
Single pion                             & ${\pm}0.21$ & ${\pm}0.27$ & 1.0   \\
Pileup                                  & ${\pm}0.35$ & ${\pm}0.31$ & 1.0  \\
Time                                    & ${\pm}0.17$ & ${\pm}0.24$ & 1.0 \\
\hline
Total JES                            & ${\pm}2.23$ & ${\pm}2.13$ & 0.9 \\
\hline
\end{tabular}
}
\end{table}

The impact of the jet energy resolution (JER) is estimated by applying
$\eta$-dependent
variations with an average of $\pm 10\%$.
The JES and JER variations are propagated to the \ETmiss. In addition,
the contribution to \ETmiss~arising from energy depositions not contained
in jets is
varied
by $\pm10\%$~\cite{jec_11}.
The uncertainty related to the  pileup modelling is determined
by propagating a $\pm 5\%$ variation~\cite{totem}
to the central value of the inelastic
cross section.
Variations in the composition of the main background processes, $\PW$+jets and
$\PZ$+jets, are conservatively evaluated by varying independently
their cross sections
by $\pm 30\%$~\cite{WWZZ8TeV, WW7TeV, ZZ7TeV}.
Additional uncertainties on the heavy flavour component in $\PW/\PZ$+jets
production are not explicitly taken into account and are assumed
to be covered by the 30\% uncertainty.
The variation of the normalization of the single top quark background by 30\%
gives a negligible contribution.
The trigger efficiency and lepton identification
correction
factors are determined with a tag-and-probe method~\cite{tag_probe}
in dilepton events and are varied within their $\pt$- and $\eta$-dependent
uncertainties.

Uncertainties from the \cPqb~tagging efficiency and mistag rate are
evaluated in the $M_3$ analysis by varying the correction factors
within their uncertainties~\cite{CMS_btag04,CMS_btag04_bis} quoted
in Section~\ref{sec:M3}.
In the $M_{\ell\cPqb}$  analysis, on the other hand,
the \cPqb~tagging efficiency for \cPqb~jets is measured
from data, using the technique described
in Refs.~\cite{CMS_btag04, CMS_btag04_bis, maes},
on the same selected event sample as that for the cross section determination,
but before \cPqb~tagging.
The $M_{\ell\cPqb}$ variable is used not only as a cross section estimator, but also
as a \cPqb~tagging discriminator.
The statistical and systematic uncertainties in the
\cPqb~tagging and mistag efficiencies are propagated to
the statistical and systematic uncertainties in the cross section
measurements.
For this reason the statistical uncertainty obtained by the $M_{\ell\cPqb}$ analysis is
larger than the one of the $M_3$ analysis.
A systematic uncertainty is assigned to the choice, based on simulation,
of the \cPqb-enriched (for $M_{\ell\cPqb}$ values below 140\GeV)
and of the \cPqb-depleted (for $M_{\ell\cPqb}$ in the range 140--240\GeV) regions,
by
shifting the windows by $\pm 30\GeV$.
Since the \cPqb~tagging efficiency and mistag rate are derived from data
and since they are
re-determined when evaluating the effect of the various systematic
uncertainties, no additional uncertainties are included.
The method is shown~\cite{CMS_btag04,CMS_btag04_bis, maes}
to be stable for different
\cPqb~tagging algorithms and working points.

Theoretical uncertainties are taken from detailed studies
performed on
simulated samples.
They include the common factorization and renormalization scales, which
are varied by a factor of 1/4 and 4 from the default value equal to the
$Q^2$
for the $\ttbar$ or $\PW/\PZ$+jet events.
The effect of the jet-parton matching threshold on $\ttbar$ and $\PW$+jets events
is studied by varying
the threshold used for matching the matrix element level to the particles
created in the parton showering by a factor of 0.5 or 2.
Uncertainties from the choice of PDF are evaluated
by using the Hessian method~\cite{hessian}
with the parameters of the CTEQ6.6 PDF set~\cite{nadolsky}.
Other PDF sets (including their uncertainties) yield very similar results.
The PDFs and their
uncertainties are determined
from a fit to collision data yielding the Hessian matrix.
Each of the 22 eigenvectors obtained by diagonalizing the matrix
is varied within its uncertainties. The differences with
respect to the nominal prediction are determined independently for each
eigenvector and are added in quadrature.
The systematic uncertainty due to the top quark $\pt$-reweighting procedure
described in Section~\ref{sec:MC} is evaluated as
the difference
with respect to the measurement obtained with the non-reweighted sample.
Only the variation due to the template shape is considered, as the correction
is meant to modify the shape only.

{\tolerance=1200
A ``signal modelling'' uncertainty is attributed to the choice
of the generators. It comprises changes in matrix element and parton
shower implementation. The effect of
the matrix element generator is evaluated by using \POWHEG
(instead of \MADGRAPH) interfaced to \PYTHIA, while
the parton shower modelling is evaluated with
\POWHEG and
\HERWIG instead of \POWHEG and \PYTHIA.
Regarding the two corresponding uncertainties, the former is
always positive and the latter is always negative.
For 7\TeV the same values determined for 8\TeV are used.
As discussed in Section~\ref{sec:comb}, the ``signal modelling'' uncertainty
is symmetrized by taking the larger of the two contributions ($\pm 4.4\%$).
}

An uncertainty of 2.6\%~\cite{lumi_2012} (2.2\%~\cite{lumi_2011})
is assigned to the determination of the 2012 (2011) integrated
luminosity.
The resulting effects from all sources are added in quadrature.
Tables~\ref{tab:syst30} and~\ref{tab:syst40} provide
an overview of the contributions to the systematic uncertainty on the
combined cross section measurements in the $M_{\ell\cPqb}$  measurements at
7 and 8\TeV.

\begin{table*}[hptb]
\centering
\topcaption{
Overview of the systematic uncertainties in the measurement of
the $\ttbar$ cross sections at 8\TeV,
both for the total
and the visible
cross sections.
For the ``signal modelling'' uncertainty
the larger between the matrix element (ME)
and parton shower (PS) uncertainties is taken,
as explained in Section~6.
The correlations assumed
for the combination of the muon and electron channels
are also given.}
\label{tab:syst30}
\renewcommand{\arraystretch}{1.2}
\begin{tabular}{lcccc} \hline
\multirow{2}{*}{Systematic uncertainty}  & \multicolumn{4}{c}{8\TeV} \\
& \multicolumn{1}{c}{$\mu$+jets (\%)}
                                  & \multicolumn{1}{c}{  $\Pe$+jets (\%)}
                                  & corr. & comb.(\%) \\
\hline
{Jet energy scale}         & ${\pm}2.2$ & ${\pm}2.1$  & 0.9 & ${\pm}2.2$ \\
{Jet energy resolution}    & ${\pm}0.8$ & ${\pm}0.9$  & 1.0 & ${\pm}0.8$ \\
{\ETmiss~unclustered energy}&${\pm}0.1$ & ${\pm}0.3$  & 1.0 & ${\pm}0.1$ \\
{Pileup}                   & ${\pm}0.5$ & ${\pm}0.4$  & 1.0 & ${\pm}0.5$ \\
{Lepton ID / Trigger eff. corrections}   & ${\pm}0.4$ & ${\pm}0.5$  & 0.0 & ${\pm}0.5$ \\
{{\cPqb} tagging method}     & ${\pm}0.3$ & ${\pm}0.7$  & 1.0 & ${\pm}0.3$ \\
{Background composition}   & ${\pm}0.2$ & ${\pm}0.3$  & 1.0 & ${\pm}0.2$ \\ \hline
{Factorization/renormalization scales}     & ${\pm}1.7$ & ${\pm}2.6$  & 1.0 & ${\pm}1.7$ \\
{ME-PS matching threshold} & ${\pm}1.3$ & ${\pm}2.3$  & 1.0 & ${\pm}1.2$ \\
{Top quark \pt-reweighting}& ${\pm}1.1$ & ${\pm}1.2$  & 1.0 & ${\pm}1.1$ \\
{Signal modelling for $\sigma_{\ttbar} (\sigma_{\ttbar}^{\vis}) $}
                           & ${\pm}4.4({\pm}2.2)$ & ${\pm}4.4({\pm}2.4)$  & 1.0 & ${\pm}4.4({\pm}2.3)$ \\
{PDF uncertainties}        & ${\pm}2.1$ & ${\pm}1.9$  & 1.0 & ${\pm}2.1$ \\ \hline

{Sum for $\sigma_{\ttbar} (\sigma_{\ttbar}^{\vis}) $}
                           & ${\pm}6.0({\pm}4.6)$ & ${\pm}6.5({\pm}5.4)$  &     & ${\pm}6.0({\pm}4.7)$ \\ \hline
{Integrated luminosity}    & ${\pm}2.6$ & ${\pm}2.6$  & 1.0 & ${\pm}2.6$ \\ \hline
{Total for $\sigma_{\ttbar} (\sigma_{\ttbar}^{\vis}) $}
                          & ${\pm}6.5({\pm}5.3)$ & ${\pm}7.0({\pm}6.0)$  &     & ${\pm}6.5({\pm}5.3)$ \\ \hline
\end{tabular}
\end{table*}

\begin{table*}[htb]
\centering
\topcaption{
Overview of the systematic uncertainties in the measurement
of the $\ttbar$ cross sections at 7\TeV,
both for the total
and the visible
cross sections.
For the ``signal modelling'' uncertainty
the larger between the matrix element (ME)
and parton shower (PS) uncertainties is taken,
as explained in Section~6.
The correlations assumed
for the combination of the muon and electron channels are also shown.
}
\label{tab:syst40}
\renewcommand{\arraystretch}{1.2}
\begin{tabular}{lcccc} \hline
\multirow{2}{*}{Systematic uncertainty}  & \multicolumn{4}{c}{7\TeV} \\
 & \multicolumn{1}{c}{$\mu$+jets (\%)}
                                  & \multicolumn{1}{c}{  e+jets (\%)}
                                  & corr. & comb.(\%) \\
\hline
{Jet energy scale}         & ${\pm}4.8$ & ${\pm}5.2$  & 0.9 & ${\pm}4.4$ \\
{Jet energy resolution}    & ${\pm}1.4$ & ${\pm}1.1$  & 1.0 & ${\pm}1.1$ \\
{\ETmiss~unclustered energy}& $<0.05$ & ${\pm}0.3$  & 1.0 & ${\pm}0.2$ \\
{Pileup}                   & ${\pm}0.4$ & ${\pm}0.6$  & 1.0 & ${\pm}0.5$ \\
{Lepton ID / Trigger eff. corrections}   & ${\pm}1.4$ & ${\pm}1.7$  & 0.0 & ${\pm}0.8$ \\
{{\cPqb} tagging method}     & ${\pm}0.5$ & ${\pm}0.6$  & 1.0 & ${\pm}0.6$ \\
{Background composition}   & ${\pm}0.5$ & ${\pm}0.4$  & 1.0 & ${\pm}0.5$ \\ \hline
{Factorization/renormalization scales}      & ${\pm}3.7$ & ${\pm}0.4$  & 1.0 & ${\pm}2.1$ \\
{ME-PS matching threshold} & ${\pm}2.0$ & ${\pm}1.7$  & 1.0 & ${\pm}1.8$ \\
{Top quark \pt-reweighting}& ${\pm}1.1$ & ${\pm}1.2$  & 1.0 & ${\pm}1.1$ \\
{Signal modelling for $\sigma_{\ttbar} (\sigma_{\ttbar}^{\vis}) $}
                           & ${\pm}4.4({\pm}2.2)$ & ${\pm}4.4({\pm}2.4)$  & 1.0 & ${\pm}4.4({\pm}2.3)$ \\
{PDF uncertainties}        & ${\pm}2.3$ & ${\pm}1.9$  & 1.0 & ${\pm}2.2$ \\ \hline

{Sum for $\sigma_{\ttbar} (\sigma_{\ttbar}^{\vis}) $}
                           & ${\pm}8.4({\pm}7.5)$ & ${\pm}7.7({\pm}6.8)$  &     & ${\pm}7.4({\pm}6.4)$ \\ \hline
{Integrated luminosity}    & ${\pm}2.2$ & ${\pm}2.2$  & 1.0 & ${\pm}2.2$ \\ \hline
{Total for $\sigma_{\ttbar} (\sigma_{\ttbar}^{\vis}) $}
                          & ${\pm}8.7({\pm}7.8)$ & ${\pm}8.0({\pm}7.1)$  &     & ${\pm}7.7({\pm}6.7)$ \\ \hline

\end{tabular}
\end{table*}

\section{Results and combination}
\label{sec:comb}

The results in the muon and electron channels,
shown in Tables~\ref{tab:results_vis} and~\ref{tab:results_tot},
are in good agreement.
The combination of the channel results is
performed using the
best linear unbiased estimator (BLUE) method~\cite{blue1, blue2, blue3}.
Asymmetric systematic uncertainties
are symmetrized for the use with BLUE  by taking half of the full range,
except for the ``signal modelling'' uncertainty, where the maximum, 4.4\%,
is taken for  $\sigma_{{\ttbar}}$.
Full correlation is assumed for all systematic uncertainties between
the two channels, except for lepton identification and trigger uncertainties,
which
are assumed to be uncorrelated.

Owing to the additional jet energy calibration from data,
a correlation coefficient of 0.9 is obtained for the overall JES uncertainty.
This correlation is determined from the correlation coefficients
in Table~\ref{tab:SYST_JES} and it is compatible with the value
inferred by comparing the combined
result with and without the additional calibration. Varying the JES
correlation coefficient between 0 and 1 has only a minor
effect on
the combined results. For example, the total cross section at 8\TeV
varies by
less than 0.5\%, and the cross section ratio varies only by
approximately 0.1\%.
A combination based on the relative statistical precision of the
two channels would also yield compatible results.
Variations of the correlations of other experimental systematic
uncertainties have negligible effect on the combined results.

The integrated luminosity and the pileup uncertainties are assumed to be fully
correlated between
channels at the same centre-of-mass energy, and uncorrelated between
7 and 8\TeV for the cross section ratio.

\subsection{Results at \texorpdfstring{$\sqrt{s}=8\TeV$}{sqrt(s) = 8 TeV}}

The visible cross section obtained from the
fit to the $M_{\ell\cPqb}$ distribution,
using \MADGRAPH signal templates
for $m_{\PQt} = 172.5\GeV$, is
\ifthenelse{\boolean{cms@external}}{
\begin{multline*}
\sigma_{\ttbar}^{\vis} (\text{combined}) \\
= 3.80 \pm 0.06\stat \pm 0.18\syst
\pm 0.10\lum\unit{pb}.
\end{multline*}
}{
\begin{equation*}
\sigma_{\ttbar}^{\vis} (\text{combined}) = 3.80 \pm 0.06\stat \pm 0.18\syst \pm 0.10\lum\unit{pb}.
\end{equation*}
}

The statistical uncertainty includes the contribution from the
simultaneous determination of the \cPqb~tagging efficiency
(see Section~\ref{sec:syst}).
There is excellent agreement with the
measurement of the visible cross section
using \POWHEG for the efficiency
within the kinematic acceptance selected by this analysis.

Using the acceptance values of Table~\ref{tab:acceptance}, the visible cross section
measurements in the electron and muon channels are first extrapolated
to the full phase space
and then combined to obtain the following
total cross section measurement
\ifthenelse{\boolean{cms@external}}{
\begin{multline*}
\sigma_{\ttbar} (\text{combined}) \\= 228.5 \pm 3.8\stat \pm 13.7\syst
\pm 6.0\lum\unit{pb}.
\end{multline*}
}{
\begin{equation*}
\sigma_{\ttbar} (\text{combined}) = 228.5 \pm 3.8\stat \pm 13.7\syst \pm 6.0\lum\unit{pb}.
\end{equation*}
}

The measurements are in good agreement with the theoretical
prediction
\begin{equation*}
\sigma^{\text{th.}}_{\ttbar}~(8\TeV)  = 252.9^{+6.4}_{-8.6} \text{(scale)} \pm 11.7 (\mathrm{PDF}{+}\alpha_S)\unit{pb}
\end{equation*}
(see Section~\ref{sec:MC}), for $m_{\PQt}=172.5\GeV$.

The BLUE combination yields the following
relative weights
of the muon
and electron channels, and their correlations, respectively.
At 8\TeV they are: $1.07$ ($1.09$), $-0.07$ ($-0.09$), with correlation
coefficient 0.88 (0.91) for the total (visible)
cross section, while at 7\TeV they are:
0.50 (0.51), 0.50 (0.49), with correlation coefficient 0.71 (0.65).
The negative
weights
of the electron channel in the combination
of the total and visible cross section at 8\TeV
depend on the choice of the JES correlation coefficient (0.9) used in
the combination. Smaller JES correlation coefficients (0.5 for the
total cross section and 0.2 for the visible cross section) would
yield positive
BLUE weights.
The negative
weights causes the combined
central value, 228.5\unit{pb},
to lie outside the interval of the two individual measurements,
as summarized in Tables~\ref{tab:results_vis} and~\ref{tab:results_tot}.

Alternatively, using \POWHEG instead of \MADGRAPH, the
combined total cross section at 8\TeV shifts by
$+8.6$\unit{pb}.
The difference, at the level of less than 4\%,
is mainly ascribed to the different acceptance for the
two generators.

All results are summarized in Tables~\ref{tab:results_vis} and~\ref{tab:results_tot}
for $m_{\PQt}=172.5\GeV$.
For \POWHEG the same relative systematic uncertainties
as determined for \MADGRAPH are used.

\begin{table*}[tbh]
 \centering
  \topcaption{
Visible cross section
measurements at $\sqrt s =$ 7 and 8\TeV
with the reference analysis $M_{\ell\cPqb}$ and the alternative analysis $M_3$ (described in Section~\ref{sec:M3}).
Results obtained for $m_{\PQt}=172.5\GeV$ with \MADGRAPH
and with \POWHEG
are shown. The uncertainties are in the order: statistical, systematic, and due to the luminosity determination.
}

\label{tab:results_vis}
\renewcommand{\arraystretch}{1.1}
    \begin{tabular}{lclc}
      \hline
Analysis & Generator & Channel & $\sigma_{\ttbar}^{\vis}$ at $\sqrt{s} = 8\TeV$ \\
\hline
\multirow{3}{*}{$M_{\ell\cPqb}$} & \multirow{3}{*}{\MADGRAPH} & $\mu$+jets     & $3.80 \pm 0.06 \pm 0.18 \pm 0.10\unit{pb}$ \\
              &           & $\Pe$+jets & $3.90 \pm 0.07 \pm 0.21 \pm 0.10\unit{pb}$ \\
              &           & Combined & $3.80 \pm 0.06 \pm 0.18 \pm 0.10\unit{pb}$ \\ \hline
$M_{\ell\cPqb}$ & \POWHEG  & Combined & $3.83 \pm 0.06 \pm 0.18 \pm 0.10\unit{pb}$ \\ \hline
\multirow{3}{*}{$M_3$}      & \multirow{3}{*}{\MADGRAPH}  & $\mu$+jets     & $3.79 \pm 0.05 \pm 0.24 \pm 0.10\unit{pb}$ \\
           &            & $\Pe$+jets & $3.75 \pm 0.04 \pm 0.26 \pm 0.10\unit{pb}$ \\
           &            & Combined & $3.78 \pm 0.04 \pm 0.25  \pm 0.10\unit{pb}$ \\ \hline
$M_3$      & \POWHEG    & Combined & $3.88 \pm 0.05 \pm 0.27 \pm 0.10\unit{pb}$ \\ \hline \hline
Analysis & Generator & Channel & $\sigma_{\ttbar}^{\vis}$ at $\sqrt{s} = 7\TeV$  \\ \hline
\multirow{3}{*}{$M_{\ell\cPqb}$ } & \multirow{3}{*}{\MADGRAPH} & $\mu$+jets     & $2.48 \pm 0.09 \pm 0.19 \pm 0.06\unit{pb}$  \\
              &           & $\Pe$+jets & $2.62 \pm 0.10 \pm 0.18 \pm 0.06\unit{pb}$  \\
              &           & Combined & $2.55 \pm 0.09 \pm 0.18 \pm 0.06\unit{pb}$  \\ \hline
    \end{tabular}
\end{table*}
\vspace{0.5cm}

\begin{table*}
 \centering
  \topcaption{
Total cross section
measurements at $\sqrt s =$ 7 and 8\TeV
with the reference analysis $M_{\ell\cPqb}$ and the alternative analysis $M_3$
(described in Section~\ref{sec:M3}).
Results obtained for $m_{\PQt}=172.5\GeV$ with  \MADGRAPH  and with \POWHEG  are shown.
The uncertainties are in the order: statistical, systematic, and due to the luminosity determination.
}
\label{tab:results_tot}
\renewcommand{\arraystretch}{1.1}
    \begin{tabular}{lclc}
      \hline
Analysis & Generator & Channel & $\sigma_{\ttbar}$ at $\sqrt{s} = 8\TeV$ \\ \hline
\multirow{3}{*}{$M_{\ell\cPqb}$} & \multirow{3}{*}{\MADGRAPH} & $\mu$+jets     & $228.9 \pm 3.4 \pm 13.7 \pm 6.0\unit{pb}$ \\
              &           & $\Pe$+jets & $234.6 \pm 3.9 \pm 15.2 \pm 6.2\unit{pb}$ \\
              &           & Combined & $228.5 \pm 3.8 \pm 13.7 \pm 6.0\unit{pb}$ \\ \hline
$M_{\ell\cPqb}$ & \POWHEG  & Combined & $237.1 \pm 3.9 \pm 14.2 \pm 6.2\unit{pb}$ \\ \hline
\multirow{3}{*}{$M_3$}      & \multirow{3}{*}{\MADGRAPH} & $\mu$+jets     & $228.7 \pm 2.6 \pm 19.0 \pm 6.0\unit{pb}$ \\
           &           & $\Pe$+jets & $225.8 \pm 2.4 \pm 19.1 \pm 5.9\unit{pb}$ \\
           &           & Combined & $227.1 \pm 2.5 \pm 19.1 \pm 6.0\unit{pb}$ \\ \hline
$M_3$      & \POWHEG   & Combined & $238.4 \pm 2.8 \pm 20.0 \pm 6.2\unit{pb}$ \\ \hline \hline
Analysis & Generator & Channel & $\sigma_{\ttbar}$ at $\sqrt{s} = 7\TeV$  \\ \hline
\multirow{3}{*}{$M_{\ell\cPqb}$} & \multirow{3}{*}{\MADGRAPH} & $\mu$+jets     &  $157.7 \pm 5.5 \pm 13.2 \pm 3.4\unit{pb}$ \\
               &           & $\Pe$+jets & $165.8 \pm 6.5 \pm 12.8 \pm 3.6\unit{pb}$ \\
               &           & Combined & $161.7 \pm 6.0  \pm 12.0 \pm 3.6\unit{pb}$  \\ \hline

    \end{tabular}
\end{table*}

\subsection{Dependence on the top quark mass at \texorpdfstring{$\sqrt{s}=8\TeV$}{sqrt(s) = 8 TeV}}

Using simulation, the dependence of the measured total cross section on the top quark mass
is determined
to be linear in the $m_{\PQt}$ range from 161.5 to 184.5\GeV.
The top quark mass value
used for the central results is 172.5\GeV.
The slope values reported in Table~\ref{tab:slope} can be used to linearly
adjust the results in the two channels to other mass values.
For $m_{\PQt}=173.3\GeV$~\cite{first_mass_comb} the adjusted
results of the two
channels yield a combined cross section value
\ifthenelse{\boolean{cms@external}}{
\begin{multline*}
\sigma_{\ttbar} (\text{combined}, m_{\PQt}=173.3\GeV) \\ =
227.4 \pm 3.8\stat
\pm 13.7\syst \pm 6.0\lum\unit{pb}.
\end{multline*}
}{
\begin{equation*}
\sigma_{\ttbar} (\text{combined}, m_{\PQt}=173.3\GeV) =
227.4 \pm 3.8\stat \pm 13.7\syst \pm 6.0\lum\unit{pb}.
\end{equation*}
}

\begin{table*}[htp]
\topcaption{Slope values
for the muon and electron channels
obtained with linear fits to the cross section values
at $\sqrt s = 8\TeV$ as a function of the
top quark mass. The \MADGRAPH generator is used.
The change in sign is due to the acceptance $A$. }
\centering
\begin{tabular}{lcc}
 \hline
 Channel & Slope (\%/\GeVns) of $\sigma^{\vis}_{\ttbar}$ & Slope (\%/\GeVns) of $\sigma_{\ttbar}$ \\
\hline
$\mu$+jets &  $+0.50 \pm 0.06$ & $-0.66 \pm 0.05$ \\
$\Pe$+jets &  $+0.30 \pm 0.04$ & $-0.94 \pm 0.05$ \\
\hline

\end{tabular}

\label{tab:slope}
\end{table*}%

\subsection{Results at \texorpdfstring{$\sqrt{s}=7\TeV$}{sqrt(s) = 7 TeV} and cross section ratio}

At $\sqrt{s} = 7\TeV$ the measured cross section, with \MADGRAPH, is
\ifthenelse{\boolean{cms@external}}{
\begin{multline*}
\sigma_{\ttbar} (\text{combined})\\ = 161.7 \pm 6.0\stat \pm 12.0\syst
\pm 3.6\lum\unit{pb}.
\end{multline*}
}{
\begin{equation*}
\sigma_{\ttbar} (\text{combined}) = 161.7 \pm 6.0\stat \pm 12.0\syst \pm 3.6\lum\unit{pb}.
\end{equation*}
}

The measurements are in good agreement with the theoretical
expectation
\begin{equation*}
\sigma^{\text{th.}}_{\ttbar}~(7\TeV) = 177.3^{+4.6}_{-6.0}\,\text{(scale)} \pm 9.0\,(\mathrm{PDF}{+}\alpha_S)\unit{pb}
\end{equation*}
at 7\TeV, for a top quark mass of 172.5\GeV.

From the measurements of the total cross section at the two centre-of-mass energies, a cross section ratio $R^{8/7}$
is determined.
In the ratio the experimental uncertainties, which are
correlated between the two
analyses (at $\sqrt{s}=7$ or 8\TeV, in each channel) cancel out, leading to
an improved precision in comparison to
the individual measurements at 7 or 8\TeV.
The ratio is first determined in the individual
muon        ($1.45 \pm 0.09$)
and electron ($1.41 \pm 0.09$)
channels and then combined.
The measured ratio
is
\begin{equation*}
R^{8/7} = 1.43 \pm 0.04\stat
               \pm 0.07\syst
               \pm 0.05\lum.
               \end{equation*}

In the combination of the ratios in the two channels
the theoretical uncertainties, and the jet-related uncertainties are assumed to be 100\% correlated,
except the
JES uncertainty, which is taken as
90\% correlated.
The other experimental uncertainties are assumed to be uncorrelated.
The expected values of the cross section ratio, for instance
$R^{8/7}_{\text{th.}} = 1.429 \pm 0.001\,\text{(scale)} \pm 0.004\,\mathrm{(PDF)}
 \pm 0.001\,\mathrm{(\alpha_s)} \pm 0.001\,(m_{\PQt})$~\cite{mitov2},
for the MSTW08 PDF set and for $m_{\PQt}=173.3\GeV$,
are in good agreement with the measurement.

\section{Alternative approach at \texorpdfstring{$\sqrt{s} = 8\TeV$}{sqrt(s) = 8 TeV} using \texorpdfstring{$M_3$}{M3}}
\label{sec:M3}

In the $M_3$ analysis similar requirements
for the selection of $\ttbar$ lepton+jets decays
are used,
with slightly different $\pt$-threshold values.
Only the differences with respect to the main selection
are summarized in the following.

At least four jets are required within $\abs{\eta} < 2.5$ and
with
$\pt > 50$, 40, 30, and 30\GeV in the muon
channel, and  $\pt > 50$, 45, 35, and 30\GeV in the electron channel.
Slightly tighter $\pt$ selection criteria are applied in the electron
channel because of the larger multijet background.
Muons are required to have transverse momentum larger than 26\GeV.
In the muon channel no explicit requirement is applied on the missing energy
in the transverse plane, while  \ETmiss has to be larger than 20\GeV
in the electron channel.

The $M_3$ analysis uses a correction factor of
$(0.95 \pm 0.02)$~\cite{CMS_btag04,CMS_btag04_bis}
to the simulated events to reproduce
the different \cPqb~tagging efficiency  in data and simulation,
and a correction factor  of $(1.11 \pm 0.01 \pm 0.12 )$~\cite{CMS_btag04,CMS_btag04_bis} to
take into account
the different probability that a light-quark or gluon jet is identified
as a \cPqb~jet.
These correction factors are determined following Refs.~\cite{CMS_btag04,CMS_btag04_bis}.
No correction factors are applied in the $M_{\ell\cPqb}$ analysis,
where these efficiencies are determined from data.

Different strategies to take into account the multijet background
are developed for the $M_{\ell\cPqb}$ and $M_3$ analyses.
In the former, this background is reduced to a negligible level thanks to
tighter selection requirements on \ETmiss and on the
transverse momenta of the third and fourth jets.
In the $M_3$ analysis, looser selection cuts are chosen and
the multijet background is considered further in the analysis.
Since MC simulation can not adequately
reproduce the shape and
normalization of multijet events,
this background is thus estimated
from data.

Selected multijet events
mostly consist
of semileptonic heavy-flavour decays and, in the electron
channel, events in which
pions in jets are misidentified as electrons.
Such events feature
lepton candidates not
coming from $\PW$~boson decays and thus not truly isolated.
The shape of the accepted multijet background is extracted from
a sideband data sample where leptons have large relative
isolation, greater than $0.17$ in the muon channel and
$0.2$ in the electron channel.
The data sample is selected such that it is rich in
multijet background and poor
in \ttbar signal and in other processes such as $\PW$+jets.
The remaining \ttbar,  $\PW$+jets and $\PZ$+jets contamination is
estimated and subtracted
using simulation. Other backgrounds, for example single top quark
production, are neglected
because of their smaller contributions.
The nominal multijet shape is taken
as the distribution measured in the sideband after subtracting
the components described above.

The template fit is performed with the  $M_3$ distribution in the range
0--1400\GeV.
One single template is used
for \ttbar events (both for the \ttbar
signal events and the other \ttbar events passing the selection requirements)
and individual templates are used
for each background process.
The $\ttbar$, single top quark, $\PW$+jets, and $\PZ$+jets templates,
used in the likelihood maximization, are taken from simulation,
while the multijet template is estimated
from data
as described above.
Because of the similarity between the single top quark
and the \ttbar templates,
the single top quark contribution is constrained by a Gaussian
distribution of 30\% width to its
expected value.
The choice of the constraint has a negligible effect on the final result.
The normalization of the signal and background processes, including the
multijet
background, is determined by the fit itself.
The muon and electron channels are combined with the BLUE method
to obtain the quoted combined result.

The measured cross section with the $M_3$ template fit is
\ifthenelse{\boolean{cms@external}}{
\begin{multline*}
\sigma_{\ttbar}(\text{combined})\\ =   227.1 \pm 2.5\stat
\pm 19.1\syst \pm 6.0\lum\unit{pb}.
\end{multline*}
}{
\begin{equation*}
\sigma_{\ttbar}(\text{combined}) =   227.1 \pm 2.5\stat \pm 19.1\syst \pm 6.0\lum\unit{pb}.
\end{equation*}
}

\begin{figure*}[tbh]
  \centering
\includegraphics[width=0.49\textwidth]{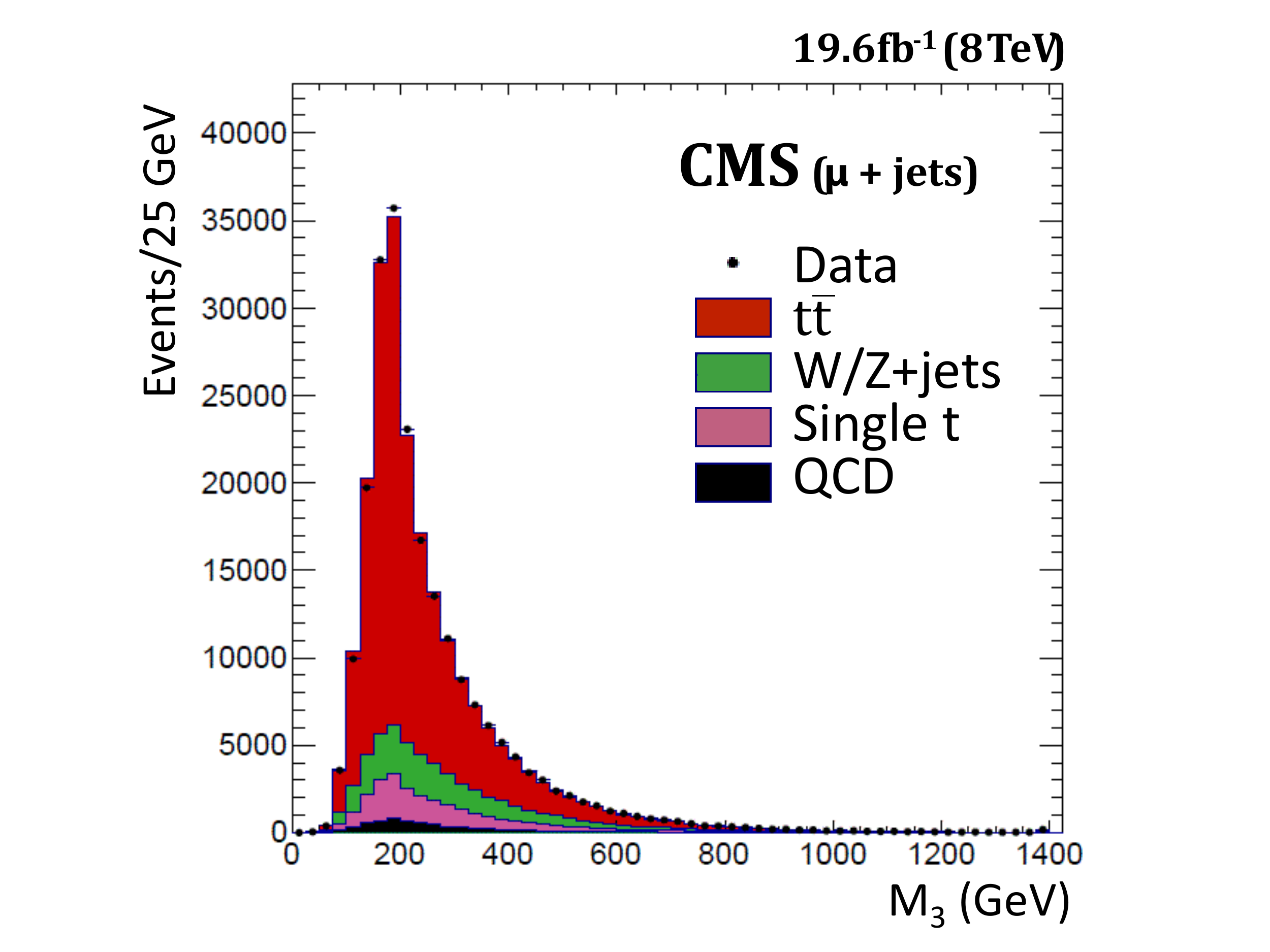}
\includegraphics[width=0.49\textwidth]{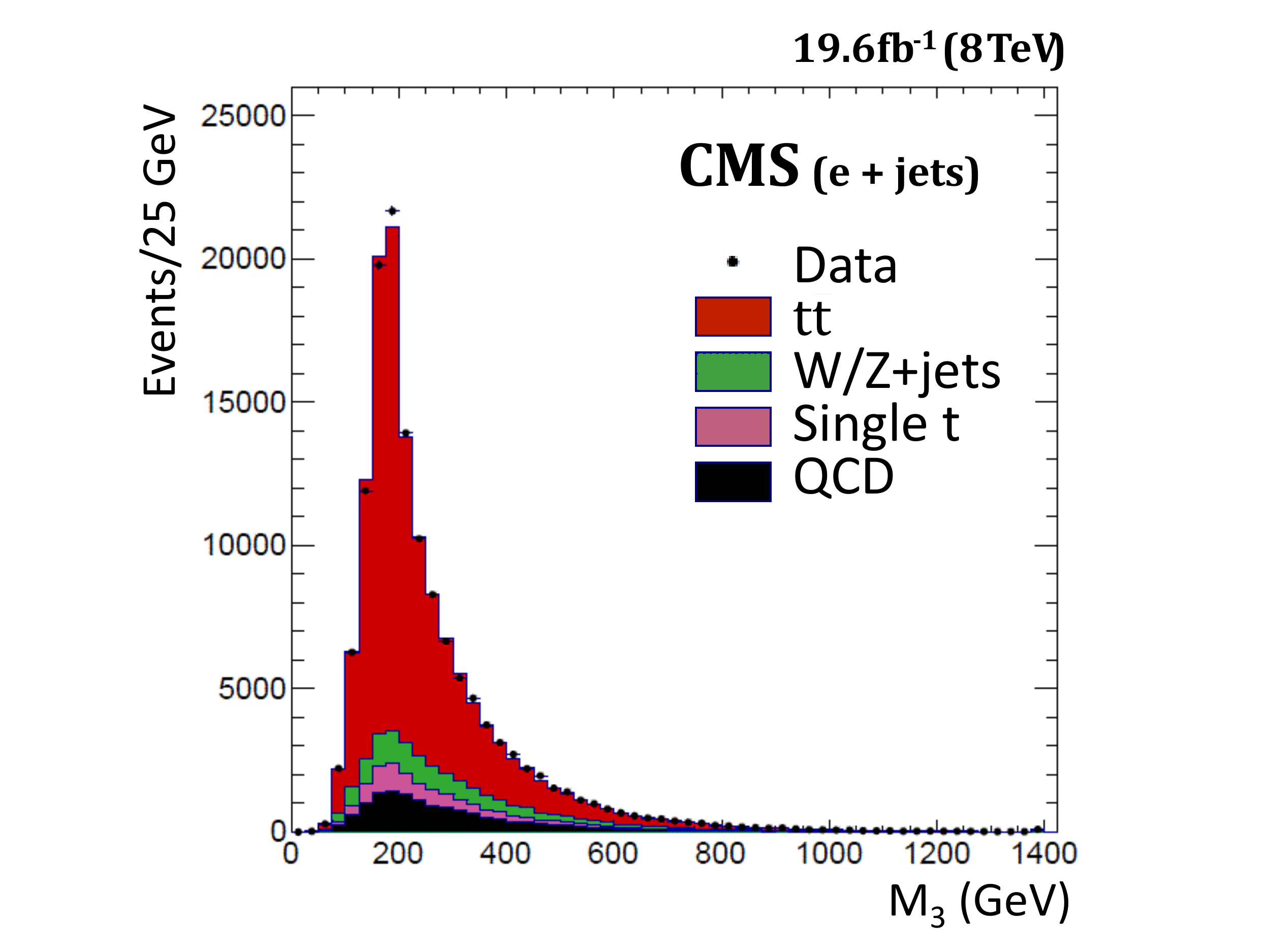}
 \caption{
Distributions of the $M_3$ mass in the 8\TeV data,
for the muon+jets (left) and electron+jets (right) channels,
rescaled to the template likelihood fit results.
The last filled bin includes the overflow.
}
\label{fig:M3fit}
\end{figure*}

{\tolerance=800
The $M_3$ distributions in the muon and electron channels are
shown in Fig.~\ref{fig:M3fit}.
Good agreement is observed between data and
the templates. The results are compatible with those of the
$M_{\ell\cPqb}$ analysis
and are summarized in Tables~\ref{tab:results_vis}
and~\ref{tab:results_tot}.
The main contributions to the systematic uncertainties
of the combined result are, in decreasing
order: signal modelling (4.4\%), factorization and renormalization scales (2.9\%),
multijet background subtraction (2.2\%),
JES (2.1\%),
PDF (1.6\%), and
\cPqb~tagging efficiency and mistag rate (1.5\%).
The uncertainty related to the multijet background subtraction is estimated
by evaluating two effects. The subtracted \ttbar, $\PW$+jets, and $\PZ$+jets
contaminations
are varied by 50\%. In addition, we assign an uncertainty to the assumption that
the $M_3$ shape does not vary in different regions of relative lepton isolation,
by repeating the analysis in six different intervals of the relative lepton isolation.
\par}

\section{Summary}
\label{sec:summary}

A measurement of the \ttbar production cross section at
$\sqrt{s}=8\TeV$
is presented, using the
data collected with the CMS detector and corresponding to
an integrated luminosity
of 19.6\fbinv.
The analysis is performed in the \ttbar
lepton+jets decay channel with one muon or electron and at least four
jets in the final state with at least one \cPqb-tagged jet.
The \ttbar cross section is extracted using a binned maximum-likelihood fit of
templates from simulated events to the data sample.
The results from the two lepton+jets channels are combined using
the BLUE method.

Techniques based on control samples in data are used
to determine the \cPqb~tagging
efficiency and to calibrate the jet energy scale. These techniques
allow for a better determination of the corresponding systematic
uncertainties, particularly for the JES, which is a
dominant source of experimental uncertainty.

In the kinematic range defined in the simulation with
exactly one muon, or electron, with $\pt > 32\GeV$  and $\abs{\eta} < 2.1 $,
one neutrino with $\pt > 40~$GeV, and at least four jets with
$\pt > 40\GeV$  and $\abs{\eta} < 2.5 $,
the measured visible $\ttbar$ cross section at
$\sqrt{s}=8\TeV$ is
$
3.80 \pm 0.06\stat \pm 0.18\syst \pm 0.10\lum\unit{pb}.
$

Using the \MADGRAPH generator for the extrapolation to the full
phase space, the total $\ttbar$ cross section at 8\TeV is
$
228.5 \pm 3.8\stat \pm 13.7\syst \pm 6.0\lum\unit{pb}.
$
The result of an alternative analysis, which makes use of the observable $M_3$,
is in good agreement with this value.

Furthermore, the analysis
performed using data at $\sqrt{s}=7\TeV$,
yields
a total cross section measurement of
$
161.7 \pm 6.0\stat
\pm 12.0\syst
\pm 3.6\lum\unit{pb}$.
The measured  cross section ratio,
where
 a number of experimental uncertainties cancel out,
is
$1.43 \pm 0.04\stat \pm 0.07\syst \pm 0.05\lum.$

All measurements are in agreement with the NNLO theoretical
predictions.

\begin{acknowledgments}
\hyphenation{Bundes-ministerium Forschungs-gemeinschaft
  Forschungs-zentren} We congratulate our colleagues in the CERN
accelerator departments for the excellent performance of the LHC and
thank the technical and administrative staffs at CERN and at other CMS
institutes for their contributions to the success of the CMS
effort. In addition, we gratefully acknowledge the computing centres
and personnel of the Worldwide LHC Computing Grid for delivering so
effectively the computing infrastructure essential to our
analyses. Finally, we acknowledge the enduring support for the
construction and operation of the LHC and the CMS detector provided by
the following funding agencies: the Austrian Federal Ministry of
Science, Research and Economy and the Austrian Science Fund; the
Belgian Fonds de la Recherche Scientifique, and Fonds voor
Wetenschappelijk Onderzoek; the Brazilian Funding Agencies (CNPq,
CAPES, FAPERJ, and FAPESP); the Bulgarian Ministry of Education and
Science; CERN; the Chinese Academy of Sciences, Ministry of Science
and Technology, and National Natural Science Foundation of China; the
Colombian Funding Agency (COLCIENCIAS); the Croatian Ministry of
Science, Education and Sport, and the Croatian Science Foundation; the
Research Promotion Foundation, Cyprus; the Ministry of Education and
Research, Estonian Research Council via IUT23-4 and IUT23-6 and
European Regional Development Fund, Estonia; the Academy of Finland,
Finnish Ministry of Education and Culture, and Helsinki Institute of
Physics; the Institut National de Physique Nucl\'eaire et de Physique
des Particules~/~CNRS, and Commissariat \`a l'\'Energie Atomique et
aux \'Energies Alternatives~/~CEA, France; the Bundesministerium f\"ur
Bildung und Forschung, Deutsche Forschungsgemeinschaft, and
Helmholtz-Gemeinschaft Deutscher Forschungszentren, Germany; the
General Secretariat for Research and Technology, Greece; the National
Scientific Research Foundation, and National Innovation Office,
Hungary; the Department of Atomic Energy and the Department of Science
and Technology, India; the Institute for Studies in Theoretical
Physics and Mathematics, Iran; the Science Foundation, Ireland; the
Istituto Nazionale di Fisica Nucleare, Italy; the Ministry of Science,
ICT and Future Planning, and National Research Foundation (NRF),
Republic of Korea; the Lithuanian Academy of Sciences; the Ministry of
Education, and University of Malaya (Malaysia); the Mexican Funding
Agencies (CINVESTAV, CONACYT, SEP, and UASLP-FAI); the Ministry of
Business, Innovation and Employment, New Zealand; the Pakistan Atomic
Energy Commission; the Ministry of Science and Higher Education and
the National Science Centre, Poland; the Funda\c{c}\~ao para a
Ci\^encia e a Tecnologia, Portugal; JINR, Dubna; the Ministry of
Education and Science of the Russian Federation, the Federal Agency of
Atomic Energy of the Russian Federation, Russian Academy of Sciences,
and the Russian Foundation for Basic Research; the Ministry of
Education, Science and Technological Development of Serbia; the
Secretar\'{\i}a de Estado de Investigaci\'on, Desarrollo e
Innovaci\'on and Programa Consolider-Ingenio 2010, Spain; the Swiss
Funding Agencies (ETH Board, ETH Zurich, PSI, SNF, UniZH, Canton
Zurich, and SER); the Ministry of Science and Technology, Taipei; the
Thailand Center of Excellence in Physics, the Institute for the
Promotion of Teaching Science and Technology of Thailand, Special Task
Force for Activating Research and the National Science and Technology
Development Agency of Thailand; the Scientific and Technical Research
Council of Turkey, and Turkish Atomic Energy Authority; the National
Academy of Sciences of Ukraine, and State Fund for Fundamental
Researches, Ukraine; the Science and Technology Facilities Council,
UK; the US Department of Energy, and the US National Science
Foundation.

Individuals have received support from the Marie-Curie programme and
the European Research Council and EPLANET (European Union); the
Leventis Foundation; the A. P. Sloan Foundation; the Alexander von
Humboldt Foundation; the Belgian Federal Science Policy Office; the
Fonds pour la Formation \`a la Recherche dans l'Industrie et dans
l'Agriculture (FRIA-Belgium); the Agentschap voor Innovatie door
Wetenschap en Technologie (IWT-Belgium); the Ministry of Education,
Youth and Sports (MEYS) of the Czech Republic; the Council of Science
and Industrial Research, India; the HOMING PLUS programme of the
Foundation for Polish Science, cofinanced from European Union,
Regional Development Fund; the Compagnia di San Paolo (Torino); the
Consorzio per la Fisica (Trieste); MIUR project 20108T4XTM (Italy);
the Thalis and Aristeia programmes cofinanced by EU-ESF and the Greek
NSRF; the National Priorities Research Program by Qatar National
Research Fund; the Rachadapisek Sompot Fund for Postdoctoral
Fellowship, Chulalongkorn University (Thailand); and the Welch
Foundation.
\end{acknowledgments}

\clearpage
\bibliography{auto_generated}

\cleardoublepage \appendix\section{The CMS Collaboration \label{app:collab}}\begin{sloppypar}\hyphenpenalty=5000\widowpenalty=500\clubpenalty=5000\textbf{Yerevan Physics Institute,  Yerevan,  Armenia}\\*[0pt]
V.~Khachatryan, A.M.~Sirunyan, A.~Tumasyan
\vskip\cmsinstskip
\textbf{Institut f\"{u}r Hochenergiephysik der OeAW,  Wien,  Austria}\\*[0pt]
W.~Adam, E.~Asilar, T.~Bergauer, J.~Brandstetter, E.~Brondolin, M.~Dragicevic, J.~Er\"{o}, M.~Flechl, M.~Friedl, R.~Fr\"{u}hwirth\cmsAuthorMark{1}, V.M.~Ghete, C.~Hartl, N.~H\"{o}rmann, J.~Hrubec, M.~Jeitler\cmsAuthorMark{1}, V.~Kn\"{u}nz, A.~K\"{o}nig, M.~Krammer\cmsAuthorMark{1}, I.~Kr\"{a}tschmer, D.~Liko, T.~Matsushita, I.~Mikulec, D.~Rabady\cmsAuthorMark{2}, B.~Rahbaran, H.~Rohringer, J.~Schieck\cmsAuthorMark{1}, R.~Sch\"{o}fbeck, J.~Strauss, W.~Treberer-Treberspurg, W.~Waltenberger, C.-E.~Wulz\cmsAuthorMark{1}
% ...
\vskip\cmsinstskip
% ...
1:~~Also at Vienna University of Technology, Vienna, Austria\\
2:~~Also at CERN, European Organization for Nuclear Research, Geneva, Switzerland\\
% ...

\end{sloppypar}
\end{document}